\documentclass[onecolumn, aps, prd, superscriptaddress, nofootinbib, notitlepage, 10pt]{revtex4-2}
\setcitestyle{numbers,square,sort&compress}
\usepackage{times, mathrsfs, amsmath, amsfonts, graphics, graphicx, cancel, color, amsthm, bbm, mathtools, amssymb, physics}
\usepackage{silence}
\usepackage{upgreek} 
\usepackage{accents}
\usepackage{dcolumn}
\usepackage{bm}
\usepackage{dsfont}
\usepackage{tikz}
\usepackage{pgfplots}
\usepgfplotslibrary{groupplots}
\pgfplotsset{compat=1.18}
\usetikzlibrary{decorations.pathreplacing}
\usetikzlibrary{arrows.meta}
\usetikzlibrary{fit}
\usepackage[hidelinks]{hyperref}
\usepackage{cleveref}
\usepackage{enumerate}
\usepackage{comment}
\usepackage{tikz}
\usepackage[normalem]{ulem}
\usepackage{xcolor}
\usetikzlibrary{arrows.meta}
\usepackage{graphicx}
\usepackage{subcaption}

\hypersetup{
colorlinks=true,
linkcolor=blue,
citecolor=red,
urlcolor=cyan
}
\let\oldparagraph\paragraph
\renewcommand{\paragraph}{\vspace{1\baselineskip}\oldparagraph}

\begin{document}

\title{Regional quantum reference frames for a massless scalar field}

\author{Giorgio Lo Giudice}
\email{giorgio.logiudice@students.uniroma2.eu}
\affiliation{Dipartimento di Fisica, Universit\`a di Roma ``Tor Vergata'' and INFN, Sezione di Roma Tor Vergata, Via della Ricerca Scientifica 1, 00133 Roma, Italy}
 \author{Davide Mattei}
 \affiliation{Dipartimento di Fisica, Universit\`a di Roma ``Tor Vergata'' and INFN, Sezione di Roma Tor Vergata, Via della Ricerca Scientifica 1, 00133 Roma, Italy}
 \author{Emanuele Panella}
 \affiliation{Dipartimento di Fisica, Universit\`a di Roma ``Tor Vergata'' and INFN, Sezione di Roma Tor Vergata, Via della Ricerca Scientifica 1, 00133 Roma, Italy}
 \author{Vittorio D'Esposito}
 \affiliation{Dipartimento di Fisica, Universit\`a di Roma ``Tor Vergata'' and INFN, Sezione di Roma Tor Vergata, Via della Ricerca Scientifica 1, 00133 Roma, Italy}
 \author{Flaminia Giacomini}
 \email{flaminia.giacomini@uniroma2.it}
 \affiliation{Dipartimento di Fisica, Universit\`a di Roma ``Tor Vergata'' and INFN, Sezione di Roma Tor Vergata, Via della Ricerca Scientifica 1, 00133 Roma, Italy}

\date{\today}

\begin{abstract}
In many physical situations, reference frames are associated to spatially-extended physical systems, whose degrees of freedom are ultimately quantum. In this work, we construct such extended quantum reference frames (QRFs) from a massless scalar field localised in a spatial region with field-shift symmetry and periodic boundary conditions. We use the lattice counterpart, a chain of coupled harmonic oscillators, to connect our construction with standard results in QRFs for systems of quantum particles. In the continuum limit of this chain, the generator of a uniform regional shift must extend into an arbitrarily thin collar of non-zero width, with implications for the localisation of the regional algebra. We derive the relational description of the field relative to a chosen region and the transformation between different regional descriptions, and show that different approaches to QRFs can be recoved based on which local observables are considered as accessible within a given region. Furthermore, we relate the ``edge mode'' connecting two regions to the so-called ``extra particle'' degree of freedom. Finally, we apply our framework to a metrological question, and find that quantum correlations between the QRF and the observable of interest enhances the precision of the measurement procedure.
\end{abstract}

\maketitle

\section{Introduction}  \label{sec:introduction}

The localisation of an event in space and time requires a reference frame structure. For such a structure to make sense operationally, it must generally extend over a finite region of space and time. For instance, a reference frame may be realised by a network of spatially distributed clocks, or by the collection of instruments in a laboratory used to measure relative distances between the production and detection of a particle. In these examples, the reference frame cannot be identified simply with the worldline of an individual observer, but must instead be understood as an extended structure in spacetime. This becomes particularly relevant in general relativity, where reference frames are often described in terms of congruences of worldlines. Crucially, the clocks and measuring instruments that realise such structures are themselves physical systems, and are ultimately quantum systems.

The notion of quantum reference frames (QRFs) extends the idea of a reference frame to the quantum domain, allowing a quantum system itself to be treated as a reference frame, from whose perspective other physical systems are described. Different approaches~\cite{AharonovKaufherr1984, BartlettRudolphSpekkens2007, GiacominiCastroRuizBrukner2019, VanrietveldeHohnGiacominiCastroRuiz2020, CastroRuizOreshkov2025RelativeSubsystems} share a common relational perspective, namely a formulation in terms of relational variables between the reference frame and the systems of interest, rather than in terms of their ``absolute'' properties or their relations to an external reference structure. To achieve this, one can either start from an external description and define a reduction procedure that extracts the relational degrees of freedom~\cite{LoveridgeMiyaderaBusch2018, VanrietveldeHohnGiacominiCastroRuiz2020, CastroRuizOreshkov2025RelativeSubsystems}, or formulate the theory relationally from the outset~\cite{GiacominiCastroRuizBrukner2019}. In both cases, one formulates a QRF transformation to switch descriptions across different QRFs. Such a QRF transformation has different properties depending on the approach and type of QRF chosen~\cite{GiacominiCastroRuizBrukner2019, DeLaHametteGalley2020, VanrietveldeHohnGiacominiCastroRuiz2020, DeLaHametteEtAl2021, CaretteGlowackiLoveridge2025, CastroRuizOreshkov2025RelativeSubsystems, GarmierHausmannCastroRuiz2025}. This perspective has recently led to a series of results extending the covariance of physical laws and symmetry principles~\cite{GiacominiCastroRuizBrukner2019, delaHamette:2021iwx} and showing that information-theoretic and physical properties depend on the choice of the QRF. These include entanglement and superposition~\cite{GiacominiCastroRuizBrukner2019}, entropy~\cite{CepollaroEtAl2025,deLaHamette2026Renyi}, thermodynamical quantities~\cite{HohnKotechaMele2023}, and open quantum system dynamics~\cite{Luppi:2026qyv}. At the interface with gravity, QRFs have been used to formulate quantum versions of Einstein's equivalence principle \cite{Hardy2020QuantumEquivalence,GiacominiBrukner2020Equivalence,GiacominiBrukner2022,CepollaroGiacomini2024EntangledClocks}, to study the relative localisation of events~\cite{CastroRuizEtAl2020QuantumClocks, Giacomini2021Spacetime, Vilasini:2025qun, KabelEtAl2025QuantumCoordinates}, and to define quantum diffeomorphisms~\cite{hardy2018construction, HametteChristodoulou, ChenGiacomini2026ReferenceFields}.

If reference frames provide an operational structure for the localisation of events in space and time, their physical realisation may require systems extending over finite spacetime regions. This raises the question of how the notion of a QRF can be generalised beyond individual quantum systems to encompass spatially extended quantum structures. Quantum fields provide a natural setting for this generalisation: their dynamical degrees of freedom can serve as physical reference structures, with the values of suitable field observables specifying where and when other observables are evaluated.
One possible way to address this generalisation uses the readings of dynamical fields to specify where and when an observable is evaluated. This idea has been developed in canonical gravity \cite{Dittrich2006,GieselThiemann2015Scalar, Thiemann:2026vaj} and in algebraic quantum field theory \cite{BrunettiFredenhagenRejzner2016, FewsterEtAl2025, Fewster:2025ijg}. In quantum cosmology, massless scalar fields provide rods and clocks, and one can study how the description of the evolution changes when a different clock is chosen \cite{HohnEtAl2012RelationalDynamics,GielenOriti2018, Hohn2019SwitchingInternalTimes, JercherMarchettiPithis2024a, JercherMarchettiPithis2024b}. In linearised quantum gravity, four dynamical scalar fields can identify the spacetime event at which a matter or a geometric observable is evaluated \cite{ChenGiacomini2026ReferenceFields}. In lattice gauge theories, quantum reference frames have shown to be related to quantum error correction and edge modes associated with the boundary of a region~\cite{AraujoRegadoHohnSartini2025, Rothlin:2026fqy, Lacambra:2026twp}. 

Here, we ask a different question: how can an extended QRF be associated with modes of a quantum field localised in a given region, and how can it be used to describe the same field from the perspective of a different region? We answer it for a massless scalar field on a compact spatial direction which enjoys a field-shift symmetry. To develop an operationally transparent picture and draw on insights from the particle-based QRF approaches, in \cref{sec:regional_quantum_reference_frames} we first discretise the theory and recognise it as a one-dimensional lattice model of coupled harmonic oscillators. The symmetry in field space reduces to a global  translational symmetry acting on the oscillator positions degrees of freedom. In such a discrete setting, the region is simply defined by a set of sites. We determine the corresponding regional algebra and the Hilbert space decomposition, both adapted to the global symmetry. 

In the continuum limit, both the reduction to a regional QRF and the transformation between descriptions relative to two different regions remain well defined (\cref{sec:quantum_reference_frame_trasformations}). However, the algebra generated by the regional reference pair is no longer contained in the algebra of observables strictly localised within the region. The reason is that the convergence of the momentum operator of the field as a generator of the regional shift requires it to have support in a region strictly largen then the one identified as a reference. This additional region, referred to as a \emph{collar} in algebraic quantum field theory~\cite{Fewster2016SplitProperty}, allows the charge profile to fall smoothly to zero. The collar can be made arbitrarily small, although it cannot be eliminated entirely.

For both the discrete and continuum case, we derive in~\cref{sec:edge_mode_qrf_approaches} an explicit relation between the edge mode, connecting different regions, and the extra particle~\cite{CastroRuizGalleyLoveridge2025InterpretingQRF,CastroRuizOreshkov2025RelativeSubsystems,SahdoCastroRuiz2026}. This sheds a different light on the connection between different QRF approaches: depending on which observables are defined as locally accessible in a given region, one recovers a specific approach to QRFs.

Finally, in~\cref{sec:energy_cost_of_precision} we apply our model to a metrology problem. Specifically, we consider a fixed excitation energy in the quantum field, and discuss measurements of relational observables. We find that the state that minimises the uncertainty on relative amplitude measurements is a squeezed state of the field. This is in line with previous results~\cite{Polzik:2014xza, Maccone2020squeezingmetrology, Fadel:2026zup}, and can be interpreted physically in terms of quantum correlations in the measurement basis helping to overcome the standard quantum limit, at the price of an increase in the uncertainty of the conjugate operators.

\section{Regional quantum reference frames}\label{sec:regional_quantum_reference_frames}

In this section, we introduce the notion of regional quantum reference frames. Our main goal is to use these as frames for the field shift symmetry of a massless scalar quantum field.  On a spatial circle of circumference $L$, the action and Hamiltonian, in units where $\hbar=1$, read
\begin{equation}
S[\Phi]= \frac{1}{2}\int_{\mathbb{R}}\mathrm{d}t\int_0^L\mathrm{d}x\,\left[(\partial_t\Phi)^2-v^2(\partial_x\Phi)^2\right] \ ,  \qquad 
H=\frac12\int_0^L \mathrm dx\,\left[\Pi(x,t)^2+v^2\left(\partial_x\Phi(x,t)\right)^2\right] \ .
\end{equation}
Since only derivatives of the field enter the action, the theory is invariant under a global field shift $\Phi(t,x)\mapsto\Phi(t,x)+\alpha$ for any constant $\alpha$ -- the freedom to choose an origin for the field amplitude is an exact symmetry of the dynamics. This shift acts only on the spatially uniform mode $\hat\Phi_0$. If the scalar has mass $M>0$, the contribution of the zero mode to the vacuum two-point function is
\begin{equation}
    \langle 0_M|\hat\Phi_0(t)\hat\Phi_0(t')|0_M\rangle
    = \frac{e^{-iM(t-t')}}{2ML}.
    \label{eq:zero_mode_two_point}
\end{equation}
Hence, it contributes to an infrared-divergent term even at finite $L$~\cite{Gibbons2015Infrared,MartinMartinezLouko2014} as the mass regulator is removed ($M\to 0$). This gives a concrete motivation for using a QRF. In the absence of an external amplitude reference, we take physical observables to be invariant under global shifts, and can read the field amplitude relative to a reference supplied by the field itself. In an equal-time difference of field amplitudes, the uniform mode contributions cancel. The resulting relative amplitude is therefore free of this infrared divergence.

The value of the field at a single spacetime point is formally an unbounded operator, meaning that extended \emph{regional} frames are needed. It is useful to consider first the corresponding theory in the discrete, namely a chain of harmonic oscillators, to leverage tools from the single particle QRF construction for the translation group. Only then we deal with the subtle issues arising from the divergent nature of QFTs.

\subsection{Regional quantum reference frames on the lattice}
\label{sec:lattice-construction}
Discretising the massless scalar field on a spatial circle of circumference $L$ gives a chain of harmonic oscillators with periodic boundary conditions~\cite{ItzyksonZuber1980,JeffersonMyers2017}. At each of the $N$ sites, the field amplitude on the lattice corresponds to the transverse displacement $\hat{\phi}_i$ of an oscillator with conjugate momentum $\hat{\pi}_i$. For lattice spacing $\Delta=L/N$, the discrete Hamiltonian reads
\begin{equation}
    \hat{H}^{\Delta}
    =
    \frac{1}{2}\sum_{i=0}^{N-1}
    \left[
        \frac{\hat{\pi}_i^2}{m}
        +
        mv^2
        \left(
            \frac{\hat{\phi}_{i+1}-\hat{\phi}_i}{\Delta}
        \right)^2
    \right]\,,
    \qquad
    \hat{\phi}_N=\hat{\phi}_0\,,
    \label{eq:lattice-hamiltonian}
\end{equation}
where $m$ is the oscillator mass (note that this does not correspond to the mass of the scalar field in the continuum, which instead arises from inserting a local harmonic potential at each site -- which we indeed lack), whilst $v$ is the long-wavelength speed of sound. Since the potential depends only on displacement differences, \cref{eq:lattice-hamiltonian} is invariant under global shift symmetry $\hat{\phi}_i\mapsto\hat{\phi}_i+\alpha$. The global shift generator is the total momentum $\hat{\pi}_T=\sum_{i=0}^{N-1}\hat{\pi}_i$, with canonical conjugate $\hat{\phi}_T=\frac{1}{N}\sum_{i=0}^{N-1}\hat{\phi}_i$. By Noether's theorem, the total momentum is a conserved charge.

\smallskip

\noindent\textbf{Reference region.} We construct a reference frame for a region $A$ containing $n_A$ sites by selecting a canonical pair of degrees of freedom, consisting of the centre-of-mass coordinate of the region and its conjugate momentum. This is a \emph{collective pair} in the sense that all sites in $A$ contribute to a single reference coordinate. This extends the well-known QRF constructions for the translation symmetry in a many-particle system, where the reference is taken to be a \emph{single} particle~\cite{AharonovKaufherr1984,BartlettRudolphSpekkens2007, GiacominiCastroRuizBrukner2019,VanrietveldeHohnGiacominiCastroRuiz2020, DeLaHametteEtAl2021}. 

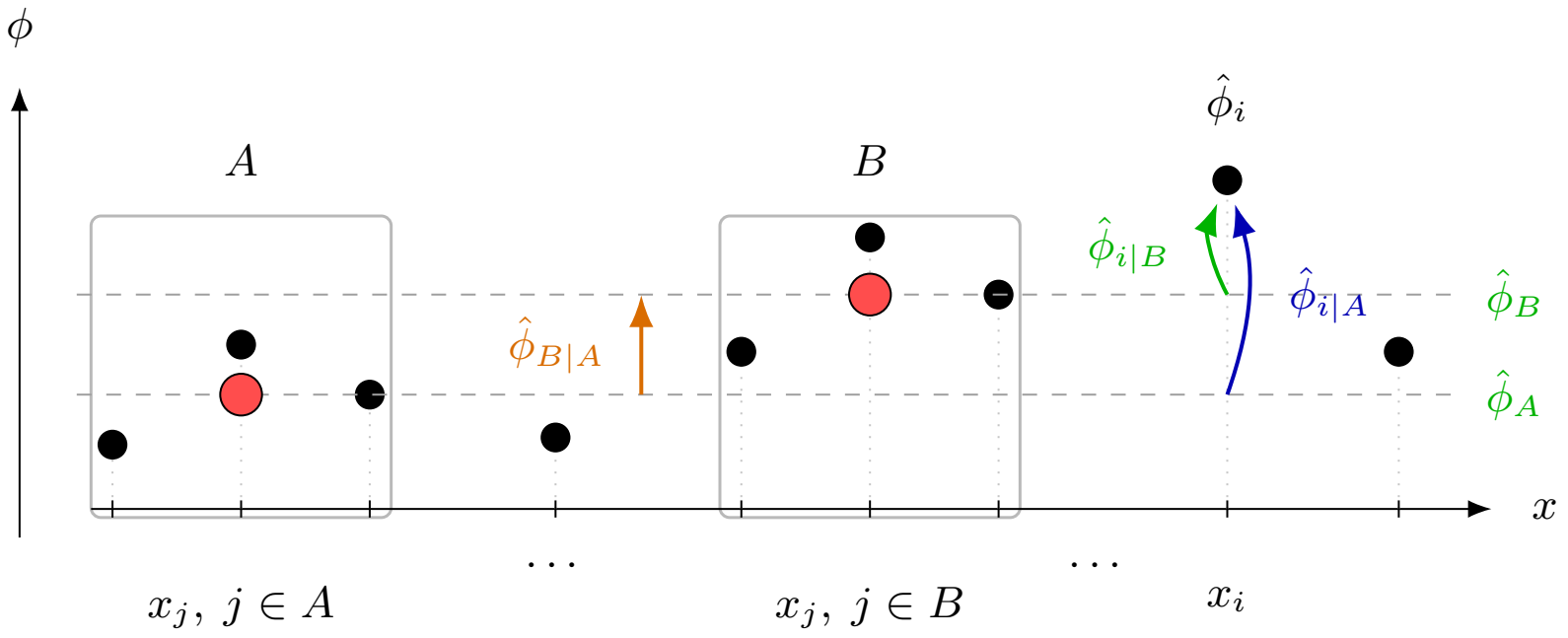
\begin{figure}[t]
\centering
\begingroup
\tikzset{
  site/.style={
    circle,draw=black,fill=black,inner sep=2pt
  },
  refsite/.style={
    circle,draw=black,fill=red!70,inner sep=3pt
  },
  regionbox/.style={
    draw=gray!55,rounded corners=2pt,line width=0.6pt
  },
  lab/.style={font=\small},
  vector/.style={->,thick,blue!70!black,shorten >=2pt},
  edgevec/.style={->,thick,orange!85!black},
  svec/.style={->,thick,green!70!black,shorten >=2pt}
}
\resizebox{0.6\linewidth}{!}{
\begin{tikzpicture}[scale=1.02,>=Latex]

\draw[regionbox] (-0.15,-0.06) rectangle (1.95,2.05);
\draw[regionbox] (4.25,-0.06) rectangle (6.35,2.05);
\node[lab,above=5pt] at (0.9,2.05) {$A$};
\node[lab,above=5pt] at (5.3,2.05) {$B$};

\draw[->,thin] (-0.15,0)--(9.65,0)
  node[right=5pt,lab] {$x$};
\draw[->,thin] (-0.65,-0.2)--(-0.65,2.95)
  node[above=5pt,lab] {$\phi$};

\foreach \name/\xx/\yy in
  {a1/0/0.45,a2/0.9/1.15,a3/1.8/0.8,
   ac1/3.1/0.5,b1/4.4/1.1,b2/5.3/1.9,
   b3/6.2/1.5,ai/7.8/2.3,aN/9.0/1.1}{
  \draw[gray!45,dotted] (\xx,0)--(\xx,\yy);
  \draw[thin] (\xx,-0.06)--(\xx,0.06);
  \node[site] (\name) at (\xx,\yy) {};
}

\node[lab,below=4pt] at (0.9,-0.3) {$x_j,\ j\in A$};
\node[lab,below=4pt] at (5.3,-0.3) {$x_j,\ j\in B$};
\node[lab,below=4pt] at (7.8,-0.3) {$x_i$};
\node[lab] at (3.1,-0.4) {$\cdots$};
\node[lab] at (6.9,-0.4) {$\cdots$};
\node[lab,above=5pt] at (ai.north) {$\hat{\phi}_i$};

\draw[gray!75,dashed] (-0.25,0.8)--(9.45,0.8);
\draw[gray!75,dashed] (-0.25,1.5)--(9.45,1.5);

\node[refsite] (XA) at (0.9,0.8) {};
\node[refsite] (XB) at (5.3,1.5) {};

\node[lab,green!70!black,anchor=west] at (9.5,0.8)
  {$\hat{\phi}_A$};
\node[lab,green!70!black,anchor=west] at (9.5,1.5)
  {$\hat{\phi}_B$};

\draw[edgevec] (3.7,0.8)--(3.7,1.5)
  node[midway,left=4pt,lab] {$\hat{\phi}_{B|A}$};
\draw[vector] (7.8,0.8) to[bend right=18]
  node[midway,right=5pt,lab] {$\hat{\phi}_{i|A}$} (ai);
\draw[svec] (7.8,1.5) to[bend left=25]
  node[midway,left=5pt,lab] {$\hat{\phi}_{i|B}$} (ai);

\end{tikzpicture}
}
\endgroup
\caption{Regional reference frames on the oscillator chain. The horizontal coordinates $x_j$ label the fixed lattice sites, while the heights $\phi_j$ depict a configuration of transverse displacements. The red markers indicate the regional mean displacements $\phi_A$ and $\phi_B$, separated by the offset $\hat{\phi}_{B|A}=\hat{\phi}_B-\hat{\phi}_A$. Switching reference from $A$ to $B$ changes the reading of a spectator site $i\notin A\cup B$ according to $\hat{\phi}_{i|B}=\hat{\phi}_{i|A}-\hat{\phi}_{B|A}$ at fixed position $x_i$.}
\label{fig:regional-qrf-discrete}
\end{figure}

To separate the reference mode from those internal to the region (bulk), we choose a different set of canonical variables adapted to the centre of mass and relative positions within the chosen region, namely
\begin{equation}
    \begin{aligned}
        \hat{\phi}_A
        &= \frac{1}{n_A}\sum_{a\in A}\hat{\phi}_a\,,
        &\qquad
        \hat{\pi}_A
        &= \sum_{a\in A}\hat{\pi}_a\,,
        \\[3pt]
        \hat{\phi}_{\mu|A}
        &= \hat{\phi}_\mu-\hat{\phi}_A\,,
        &
        \hat{\pi}_{\mu|A}
        &= \hat{\pi}_\mu-\hat{\pi}_{a_*}\,,
    \end{aligned}
    \label{eq:discrete-regional-cm}
\end{equation}
where $a_*\in A$ is any fixed site and $\mu\in A\setminus\{a_*\}$. The first row is the collective reference pair. The second row gives a canonical chart for the bulk of $A$. The displacement of an oscillator $i$ inside or outside region $A$ is expressed relationally to $A$ as $\hat{\phi}_{i|A}=\hat{\phi}_i -\hat{\phi}_A$, as illustrated in \Cref{fig:regional-qrf-discrete}. Outside the region, the conjugate variable to $\hat{\phi}_{i|A}$ is simply $\hat{\pi}_i$ itself. On the lattice, the collective reference pair, consisting of the regional centre of mass and its conjugate momentum, is supported in $A$. As we shall see, this does not hold in the continuum limit – and this affects the definition of the regional algebra of the QFT in a crucial way.

\smallskip  \noindent\textbf{Hilbert space decomposition.} We now turn our attention to the physical regional algebra of observables $\mathcal{A}_{\mathrm{phys,A}}$, from which a natural decomposition of the Hilbert space can be derived. 

The total physical algebra $\mathcal{A}_{\mathrm{phys}}$ is generated by the displacement differences $\hat{\phi}_i-\hat{\phi}_j$ and the individual momenta $\hat{\pi}_i$. The regional algebra, without the “physical” restriction, is instead generated by the site operators $\hat{\phi}_a$ and $\hat{\pi}_a$ with $a\in A$. Their intersection, $\mathcal{A}_{\mathrm{phys},A} =\mathcal{A}_{\mathrm{phys}}\cap\mathcal{A}_A$, therefore contains precisely the shift-invariant observables accessible within $A$. The variables in \cref{eq:discrete-regional-cm} make the structure of this intersection explicit. A global shift leaves all bulk variables and the regional momentum $\hat{\pi}_A$ invariant, while translating the regional mean $\hat{\phi}_A$. Hence $\mathcal{A}_{\mathrm{phys},A}$ is generated by the bulk algebra, denoted as $\mathcal{A}_A^b$, together with $\hat{\pi}_A$. Since the regional momentum commutes with every bulk variable, its spectral projections generate the centre of $\mathcal{A}_{\mathrm{phys},A}$. The value of $\pi_A$ thus labels sectors that invariant observables within $A$ can distinguish but cannot connect.

Having identified the regional algebra and its charge sectors, we can now write the corresponding Hilbert-space decomposition. For two complementary regions $A$ and $\bar A$, the canonical variables separate the collective and bulk modes on each side. In a basis of eigenvectors adapted to the charges, it is easy to see that, at total charge $\pi_T$, a charge $\pi_A$ in $A$ must be paired with charge $\pi_T-\pi_A$ in $\bar A$ \cite{Bianchi:2024NonAbelianSREE,GarmierHausmannCastroRuiz2025}. Hence, the decomposition of the Hilbert space in the two regions is
\begin{equation}
    \begin{aligned}
        \mathcal{H}_{\mathrm{kin}}
        &\simeq
        \bigl(
            \mathcal{H}_{A}\otimes\mathcal{H}_A^b
        \bigr)
        \otimes
        \bigl(
            \mathcal{H}_{\bar A}\otimes\mathcal{H}_{\bar A}^b
        \bigr)
        \\[3pt]
        &\simeq
        \int_{\mathbb R}^{\oplus}\!\mathrm{d}\pi_T\;
        \mathcal{H}_G^{(\pi_T)}
        \otimes
        \left[
            \int_{\mathbb R}^{\oplus}\!\mathrm{d}\pi_A\;
            \mathcal{H}_{\mathrm{phys},A}^{(\pi_A)}
            \otimes
            \mathcal{H}_{\mathrm{phys},\bar A}^{(\pi_T-\pi_A)}
        \right]\,.
    \end{aligned}
    \label{eq:lattice-hilbert-space-factorization-kinematical}
\end{equation}
In the second line $\mathcal{H}_{\mathrm{phys}} \simeq \int_{\mathbb R}^{\oplus}\!\mathrm{d}\pi_A\ \mathcal{H}_{\mathrm{phys},A}^{(\pi_A)} \otimes \mathcal{H}_{\mathrm{phys},\bar A}^{(\pi_T-\pi_A)} $. Note that $\mathcal{H}_{\bar A}^b$ contains field variables relativised with respect to $\bar{A}$, i.e. the pairs $(\hat{\phi}_{\nu|\bar{A}},\hat{\pi}_{\nu|\bar{A}})$ with the site $\nu$ in $\bar{A}$. Here $\mathcal{H}_G^{(\pi_T)}$ carries the one-dimensional representation of global shifts with total momentum $\pi_T$. The two regional factors in the inner integral carry the invariant regional algebras at the indicated charges. Explicit computations showing the Hilbert space decomposition induced by the regional algebras can be found in~\cref{app:subsystem-algebras}.

Let us pause to connect our discussion with the QRF literature, specifically to the different ways invariance under a symmetry group can be imposed. The outer direct integral in \cref{eq:lattice-hilbert-space-factorization-kinematical} provides the setting for the distinction between weak and strong implementations of the symmetry~\cite{BartlettRudolphSpekkens2007,VanrietveldeHohnGiacominiCastroRuiz2020,HohnKrummMueller2022,AugustinTotalCharge}. For the Hamiltonian in \cref{eq:lattice-hamiltonian}, $\hat{\pi}_T$ generates a physical symmetry and is conserved. In the weak approach, we restrict to observables commuting with $\hat{\pi}_T$, while retaining all total-charge sectors. The total charge is then central in the invariant algebra, whose observables preserve each sector and cannot detect coherences between different sectors.

A stronger restriction selects one total-charge sector, formally imposing $(\hat{\pi}_T-\pi_T\mathbb I)|\Psi\rangle=0$. In a constrained formulation, $\hat{\pi}_T$ generates global shifts treated as gauge transformations.\footnote{Such a formulation can be obtained by reformulating the Hamiltonian to be invariant under arbitrary time-dependent uniform shifts
$\phi_i(t)\mapsto\phi_i(t)+\alpha(t)$; see \cite{VanrietveldeHohnGiacominiCastroRuiz2020}.} In either description, within a sector of total charge $\pi_T$, the regional charge labels satisfy $\pi_{\bar A}=\pi_T-\pi_A$, although neither regional charge is fixed individually. Eq. \eqref{eq:lattice-hilbert-space-factorization-kinematical} then reduces to
\begin{equation}
    \mathcal{H}_{\mathrm{phys}}^{(\pi_T)}
    \simeq
    \int_{\mathbb R}^{\oplus}\!\mathrm{d}\pi_A\;
    \mathcal{H}_{\mathrm{phys},A}^{(\pi_A)}
    \otimes
    \mathcal{H}_{\mathrm{phys},\bar A}^{(\pi_T-\pi_A)}\,.
    \label{eq:phys-edge-discrete}
\end{equation}
We consider this fixed-charge description in~\cref{sec:edge_mode_qrf_approaches}, where we connect QRFs and regional observable algebras with the role of edge modes in local gauge theories. This additional restriction is not required for the construction of regional references and relative observables developed here.

\subsection{Continuum limit} \label{sec:continuum_limit}
In the double scaling limit $\Delta\to 0$ and $N\to \infty$, with $L$ kept constant, the lattice theory recovers the original massless scalar field (a non-singular continuum limit requires also  $m/\Delta$  and   $v$ to remain fixed, for simplicity we take $m/\Delta=1$):
\begin{equation}
\phi_i(t)\to\Phi(x,t)\,, \qquad \frac{\pi_i(t)}{\Delta}\to\Pi(x,t)\,.
\end{equation}
Let us see the rest of the dictionary from the lattice to the continuum explicitly. For fixed physical wave number $k_n=2\pi n/L$, the lattice dispersion satisfies
\begin{equation}
\omega_n = \frac{2v}{\Delta} \sin \left(\frac{k_n\Delta}{2}\right) \to vk_n \qquad (k>0)\,.
\end{equation}
Thus the continuum theory has dispersion $\omega(k)=v|k|$ and equation of motion
\begin{equation}
(\partial_t^2-v^2\partial_x^2)\Phi(x,t)=0\,.
\end{equation}
Finally, taking $\Delta\to0$ at fixed $L=N\Delta$, the global translation generator becomes
\begin{equation}
 \Pi_0
 = \lim_{\Delta\to0}\sum_{i=0}^{N-1}
   \Delta\,\frac{p_i}{\Delta}
 = \int_0^L \dd{y}\,\Pi(y)\,.
\end{equation}

To isolate the degree of freedom affected by the shift symmetry, we separate the zero mode from the rest of the Fourier modes. At fixed time, we write
\begin{equation}
\begin{aligned}
 \hat\Phi(x)
 &= \hat\Phi_0+\hat\Phi_{\mathrm{osc}}(x)\,,
 &\qquad
 &\hat\Phi_0
 = \frac{1}{L}\int_0^L \dd{x}\,\hat\Phi(x)\,,\\
 \hat\Pi(x)
 &= \frac{\hat \Pi_0}{L}+\hat\Pi_{\mathrm{osc}}(x)\,,
 &\qquad
 &[\hat\Phi_0,\hat \Pi_0]
 = i\hbar\,.
\end{aligned}
\end{equation}
The oscillating modes have zero spatial average. For the real scalar field considered here, it is useful to pair the Fourier components at opposite wave numbers and use the orthonormal basis $e_{n,c}(x)=\sqrt{2/L}\cos(k_n x)$ and $e_{n,s}(x)=\sqrt{2/L}\sin(k_n x)$ for $n\geq 1$. Writing $j=(n,c)$ or $j=(n,s)$, we have
\begin{equation}
    \begin{aligned}
        \hat\Phi_{\mathrm{osc}}(x)
        &= \sum_j e_j(x)\,\hat q_j\,,
        &\qquad
        \hat q_j
        &= \int_0^L e_j(x)\,\hat\Phi_{\mathrm{osc}}(x)\,\mathrm{d}x\,,
        \\[4pt]
        \hat\Pi_{\mathrm{osc}}(x)
        &= \sum_j e_j(x)\,\hat p_j\,,
        &
        \hat p_j
        &= \int_0^L e_j(x)\,\hat\Pi_{\mathrm{osc}}(x)\,\mathrm{d}x\,.
    \end{aligned}
    \label{eq:real-fourier-mode-expansion}
\end{equation}
Orthonormality gives $[\hat q_j,\hat p_{j^\prime}]=i\hbar\delta_{jj^\prime}$. We use this decomposition explicitly in \cref{sec:energy_cost_of_precision}.

The pair $(\hat\Phi_0,\hat\Pi_0)$ is the continuum counterpart of $(\hat\phi_T,\hat\pi_T)$ on the lattice. To make its physical role explicit, let $\hat U(a)=e^{-ia\hat\Pi_0/\hbar}$ and denote the global field shift by $\mathcal T_a[\hat O]\coloneqq\hat U(a)^\dagger\hat O\hat U(a)$. Its action on the field decomposition is
\begin{equation}
    \mathcal T_a[\hat\Phi_0]=\hat\Phi_0+a\mathds 1\,,
    \qquad
    \mathcal T_a[\hat\Phi_{\mathrm{osc}}(x)]
    =\hat\Phi_{\mathrm{osc}}(x)\,.
    \label{eq:global-field-shift}
\end{equation}
Thus, a constant global change in the amplitude shifts the zero mode by $a$ while leaving every non-zero mode unchanged. In the absence of an external amplitude reference, configurations related by this shift describe the same physics. For example, the mean amplitudes of two regions both shift by $a$, so their difference is independent of the choice of origin.

Analogously to what happens on the lattice, identifying configurations related by $\mathcal T_a$ removes the absolute field amplitude $\hat\Phi_0$ from the physical description. The total charge $\hat\Pi_0$ remains a central observable and acts as $Q\mathds{1}$ in each charge sector. Thus, the global pair effectively no longer constitutes a canonical degree of freedom.

\subsection{Regional quantum reference frames for fields}\label{sec:regional_construction_conditions}

Since quantum fields are operator-valued distributions, they formally require smearing~\cite{ItzyksonZuber1980}; we use the notation
\begin{equation}
 \hat\Phi(r)=\int_0^L \dd{x}\,r(x)\hat\Phi(x)\,,\qquad \hat\Pi(g)=\int_0^L \dd{x}\,g(x)\hat\Pi(x)\,.
\end{equation}
We would like to extend the construction introduced for the lattice model to the field case. On the lattice, our regional reference frame was defined by the collective canonical pair $(\phi_A,\pi_A)$. A naive extension of this construction suggests using a sharp window for the regional field average and its conjugate charge
\begin{equation}
\phi_A\to \Phi(\frac{\chi_A}{l_A})=\frac{1}{l_A} \int_0^L \dd{x}\, \chi_A(x)\Phi(x)\qquad   \pi_A \to \Pi(\chi_A)=\int_0^L \dd{x}\, \chi_A(x)\Pi(x)\,,\label{eq:window_function_region}
\end{equation}
where the window function $\chi_A(x)=1$ if $x \in A$, otherwise $\chi_A(x)=0$. However, this choice yields ill-defined observables in the continuum. The reason is that the squared vacuum norm, although finite for $\hat\Phi_{\mathrm{osc}}(\chi_A)$, diverges logarithmically for $\hat\Pi_{\mathrm{osc}}(\chi_A)$~\cite{Agullo2023}. The derivation is given in \cref{app:regularity_of_smeared_operators}. Hence, the regional smearing function needs to be smooth.

To obtain a valid regional QRF description for the massless scalar field, we adopt the following prescription. Each reference is specified by a field average $\hat\Phi(r_\alpha)\coloneqq\Phi_\alpha$ and its conjugate charge $\hat\Pi(g_\alpha)\coloneqq\Pi_\alpha$ with $\alpha=1, \dots, m$ and $m\in\mathbb{N},\hspace{1mm} $, with $m$ finite. The subscript $\alpha$ refers to the region considered, $A_\alpha$, which is defined as a closed arc on the circle. We choose real smearing profiles $r_\alpha$ and $g_\alpha$ satisfying the following conditions:

\begin{enumerate}
\item \textbf{Regionality.} 
The smearing profile $r_\alpha$ for the reference variable $\Phi_\alpha$ needs to have compact support in the corresponding region $A_\alpha$
\begin{equation}
    \operatorname{supp} r_\alpha = A_\alpha\,.
\end{equation}
As we shall see, one cannot require the same for the charge smearing profile $g_\alpha$.
    \item \textbf{Duality.}
    Each translation should only act non-trivially within its corresponding region. Hence, each regional charge $\hat{\Pi}_\beta$ should commute with $\hat{\Phi}_\alpha$ for $\alpha \neq \beta$. This leads to:
    \begin{equation}
        \int_0^L \dd{x}\,r_\alpha(x)g_\beta(x)
        = \delta_{\alpha\beta}\,.
    \end{equation}
\item \textbf{Partition of unity.}
    The individual regional charges must sum to the generator of uniform
    field shifts
    \begin{equation}
        \sum_\gamma g_\gamma(x)=1.
    \end{equation}  
\item \textbf{Positivity.}
The smearing profiles should not be negative,
\begin{equation}
    r_\alpha(x)\geq0\,,
    \qquad
    g_\alpha(x)\geq0\,.
\end{equation}
 Positivity of $r_\alpha$ makes $\hat\Phi_\alpha$ a weighted average of the field, mirroring the lattice construction. Positivity of $g_\alpha$ preserves the interpretation of $\hat\Pi_\alpha$ as a smooth counterpart of the total momentum of a group of lattice sites.
    \item \textbf{Regularity.}
    The smeared observables must have finite oscillator vacuum norms. Smooth periodic profiles ensure that
    \begin{equation}
    \bigl\|\hat\Phi_{\mathrm{osc}}(r_\alpha)
        \ket{0_{\mathrm{osc}}}\bigr\|^2 < \infty\,, \qquad
        \bigl\|\hat\Pi_{\mathrm{osc}}(g_\alpha)
        \ket{0_{\mathrm{osc}}}\bigr\|^2 < \infty\,.
    \end{equation}

\end{enumerate}

These conditions together imply a crucial, further property -- the existence of a buffer space between the regions. Indeeed, the conjugate charge profile $g_\alpha$ needs to equal unity throughout $A_\alpha$, including its endpoints. Smoothness therefore requires its support to extend beyond both endpoints. We define 
\begin{equation}
    A_\alpha^+ \coloneqq \operatorname{supp} g_\alpha\,,
    \qquad
    \mathcal C_\alpha \coloneqq A_\alpha^+ \setminus A_\alpha\,,
\end{equation}
where $\mathcal C_\alpha$ is typically referred to as “collar” \cite{Fewster2016SplitProperty} and accommodates the smooth transition of the charge profile outside $A_\alpha$. We denote the size of such a collar region $\ell$; our conditions do not fix $\ell$ given a choice of a single $r_{\alpha}$. 
See \cref{fig:single-region-sharp-limit} for an illustration of the smearing functions $r_\alpha$ and $g_\alpha$.
\pgfmathdeclarefunction{qrfProfileStep}{1}{\pgfmathparse{exp(-1/(#1))/(exp(-1/(#1))+exp(-1/(1-(#1))))}}
\newcommand{\qrfPlotAverage}[4]{\draw[#4,very thick,line join=round, domain=-0.999:0.999,samples=101,variable=\u] plot ({(#1)+(#2)*\u/2},
{(#3)*exp(-1/(1-\u*\u))/0.221996908084});}
\newcommand{\qrfPlotCharge}[5]{ \draw[#5,very thick] ({(#1)-(#2)/2},#4) -- ({(#1)+(#2)/2},#4); \foreach \side in {-1,1}{\draw[#5,very thick,line join=round, domain=0.001:0.999,samples=61,variable=\t] plot ({(#1)+\side*((#2)/2+(#3)*\t)}, {(#4)*(1-qrfProfileStep(\t))});}}

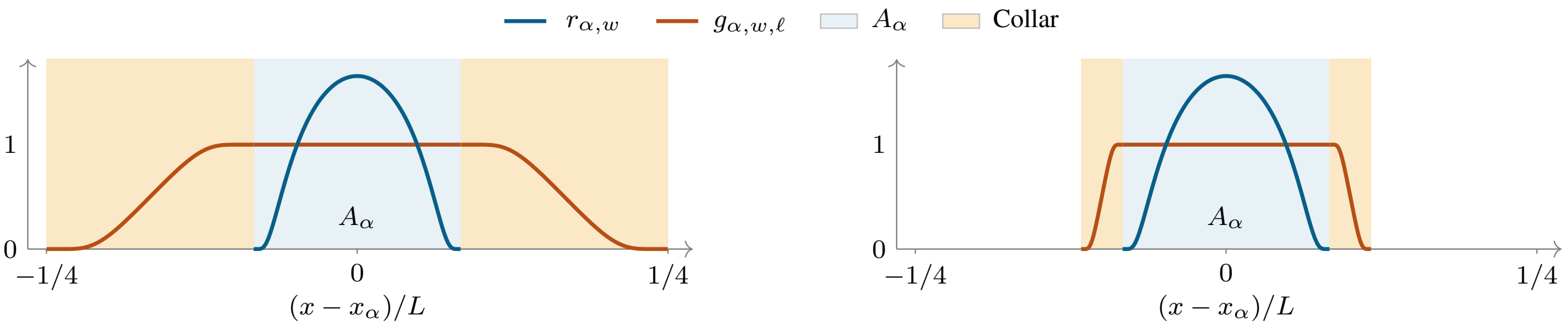
\begin{figure}[htbp]
\centering
\begingroup

\definecolor{qrfField}{HTML}{075F88}
\definecolor{qrfCharge}{HTML}{B64F16}
\definecolor{qrfArc}{HTML}{E8F1F6}
\definecolor{qrfCollar}{HTML}{FAE8C6}

\newcommand{\qrfLegendLine}[1]{\tikz[baseline=-0.5ex]{
  \draw[#1,very thick] (0,0) -- (1.6em,0);
}}
\newcommand{\qrfLegendBox}[1]{\tikz[baseline=-0.5ex]{
  \path[fill=#1,draw=black!25,thin]
    (0,-0.6ex) rectangle (1.4em,0.6ex);
}}

{\footnotesize
  \qrfLegendLine{qrfField}\enspace $\,r_{\alpha,w}$
  \quad
  \qrfLegendLine{qrfCharge}\enspace $g_{\alpha,w,\ell}$
  \quad
  \qrfLegendBox{qrfArc}\enspace $A_\alpha$
  \quad
  \qrfLegendBox{qrfCollar}\enspace Collar
  \par
}
\medskip

\newcommand{\qrfSingleArc}[1]{\begin{minipage}[t]{0.48\linewidth}
  \centering
  \begin{tikzpicture}[x=1.55\linewidth,y=1.12cm,font=\footnotesize]
    \pgfmathsetmacro{\wid}{1/6}
    \pgfmathsetmacro{\halfarc}{\wid/2}
    \pgfmathsetmacro{\ellRatio}{1/#1}

    \fill[qrfArc]
      (-\halfarc,0) rectangle (\halfarc,1.82);
    \fill[qrfCollar]
      ({-\halfarc-\ellRatio},0) rectangle (-\halfarc,1.82);
    \fill[qrfCollar]
      (\halfarc,0) rectangle ({\halfarc+\ellRatio},1.82);

    \draw[black!50,->] (-0.265,0) -- (0.27,0);
    \draw[black!50,->] (-0.265,0) -- (-0.265,1.82);
    \foreach \y in {0,1}{
      \node[left] at (-0.265,\y) {$\y$};
    }
    \foreach \x/\lab in {-0.25/{-1/4},0/0,0.25/{1/4}}{
      \draw[black!50] (\x,0) -- (\x,-0.045)
        node[below,text=black] {$\lab$};
    }

    \qrfPlotCharge{0}{\wid}{\ellRatio}{1}{qrfCharge}
    \qrfPlotAverage{0}{\wid}{1}{qrfField}

    \node at (0,0.30) {$A_\alpha$};
    \node[below] at (0,-0.34) {$(x-x_\alpha)/L$};
  \end{tikzpicture}
\end{minipage}}

\qrfSingleArc{6}\hfill\qrfSingleArc{30}\par

\caption{The collar width can be varied independently of the regional averaging profile. Both panels show the same reference region $A_\alpha$ (light blue), centred at $x_\alpha$ and of fixed width $w=L/6$. The blue curve $r_{\alpha,w}$ is identical in both panels and has support throughout the region. Each yellow strip has width $\ell=L/6$ on the left and $\ell=L/30$ on the right. As the collar narrows, the orange charge profile $g_{\alpha,w,\ell}$ approaches the characteristic function $\chi_{A_\alpha}$.}
\label{fig:single-region-sharp-limit}

\endgroup
\end{figure}

\smallskip  \noindent\textbf{Choice of the smearing profiles.}
We now give an explicit form of regional smearing profiles that satisfy our conditions (more details about the smearing profiles can be found in \cref{app:explicit_regional_profiles}). Let $x_\alpha^\pm=x_\alpha\pm w/2$ denote the endpoints of the region of width $w$.  For the field average, we choose a smooth profile that gives greater weight to points near the centre of the region. Writing $u_{\alpha,w}(x)=2(x-x_\alpha)/w$, we take
\begin{equation}
\begin{aligned}
r_{\alpha,w}(x) &= \frac{1}{Z_w} 
\begin{cases}
\displaystyle \exp\!\left[-\frac{1}{1-u_{\alpha,w}(x)^2}\right]\,,
& |u_{\alpha,w}(x)|<1,\\[6pt]
0\,, & |u_{\alpha,w}(x)|\geq1\,,
\end{cases}\\[4pt]
Z_w&=\frac{w}{2}\int_{-1}^{1} \exp\!\left[-\frac{1}{1-u^2}\right]\mathrm{d}u\,.
\end{aligned}
\label{eq:regional-reference-profile}
\end{equation}
This profile has unit integral and is supported exactly in $A_\alpha$. Its shape as a function of $u_{\alpha,w}$ specifies the relative weight of different parts of the region. 
 For the charge profile, the regional conditions discussed in the previous section require $g_{\alpha,w,\ell}=1$ throughout $A_\alpha$. To describe its transition to zero outside the region, we use the smooth step
\begin{equation}
S(t)=
\begin{cases}
0\,, & t\leq0\,,\\[4pt]
\displaystyle
\frac{e^{-1/t}}{e^{-1/t}+e^{-1/(1-t)}}\,,
& 0<t<1,\\[8pt]
1\,, & t\geq1 \,.
\end{cases}
\label{eq:regional-smooth-step}
\end{equation}
A rising step across the left collar and a falling step across the right collar give
\begin{equation}
g_{\alpha,w,\ell}(x)= S\!\left(1+\frac{x-x_-}{\ell}\right)S\!\left(1+\frac{x_+-x}{\ell}\right)\,.
    \label{eq:regional-charge-profile}
\end{equation}
In the previous equation, the first factor rises from zero to one across the left portion of the collar, $[x_- - \ell,x_-]$, and the second falls from one to zero across the right portion, $[x_+,x_+ + \ell]$. Both factors equal one throughout $A_\alpha$, so their product has the plateau required by the regional conditions above. Since $r_{\alpha,w}$ is normalised and supported on $A_\alpha$, the duality relation $\int_0^L r_{\alpha,w}(x)\,g_{\alpha,w,\ell}(x)\,\mathrm{d}x=1$ holds for every admissible $\ell>0$. 

\begin{figure}[htbp]
\centering
\begingroup
\definecolor{qrfA}{RGB}{36,88,166}
\definecolor{qrfB}{HTML}{176B42}
\definecolor{qrfC}{RGB}{156,54,142}
\definecolor{qrfCollar}{HTML}{FAE8C6}
\newcommand{\qrfThreeArcs}[2]{\begin{minipage}[t]{0.48\linewidth}
\centering
{\small $\ell=L/#1,\quad w=#2$\par}
\smallskip
\begin{tikzpicture}[x=0.85\linewidth,y=0.95cm,font=\footnotesize]
\pgfmathsetmacro{\gap}{1/#1}
\pgfmathsetmacro{\wid}{1/3-\gap}
\foreach \base/\isfield in {0/1,-2.55/0}{\begin{scope}[shift={(0,\base)}]
\fill[qrfCollar] (0,0) rectangle (1,1.65);
\foreach \c/\col/\edgecol in
{0.166667/qrfA/black!25,0.5/qrfB/qrfB!50, 0.833333/qrfC/black!25}{
\fill[\col!10]({\c-\wid/2},0) rectangle ({\c+\wid/2},1.65);\foreach \side in {-1,1}{\draw[\edgecol,thin]({\c+\side*\wid/2},0) -- ({\c+\side*\wid/2},1.65);}}
\begin{scope}
\clip (0,-0.015) rectangle (1,1.65);
\ifnum\isfield=1\relax
\foreach \c/\col in {0.166667/qrfA,0.5/qrfB,0.833333/qrfC}{\qrfPlotAverage{\c}{\wid}{0.13/\wid}{\col}}
\else
\foreach \c/\col in{-0.166667/qrfC,0.166667/qrfA,0.5/qrfB,0.833333/qrfC,1.166667/qrfA}{\qrfPlotCharge{\c}{\wid}{\gap}{1.30}{\col}}
\fi
\end{scope}
\draw[black!65,->] (0,0) -- (1.035,0);
\draw[black!65,->] (0,0) -- (0,1.72);
\node[left] at (0,0) {$0$};
\ifnum\isfield=1\relax
\node[left] at (0,0.65) {$5$};
\node[left] at (0,1.30) {$10$};
\node[above right] at (0,1.73) {$Lr_\gamma(x)$};
\foreach \c in {0.166667,0.5}{\draw[black!55,thin]({\c+\wid/2},1.68) -- ({\c+\wid/2},1.75)-- ({\c+\wid/2+\gap},1.75)-- ({\c+\wid/2+\gap},1.68);\node[above] at({\c+\wid/2+\gap/2},1.75) {$\ell$};}
\else
\node[left] at (0,0.65) {$\frac12$};
\node[left] at (0,1.30) {$1$};
\node[above right] at (0,1.70) {$g_\gamma(x)$};
\fi
\foreach \c/\col/\lab in {0.166667/qrfA/A,0.5/qrfB/B,0.833333/qrfC/C}{\ifnum\isfield=1\relax\node[text=\col] at (\c,1.49) {$r_{\lab}$};
\else
\node[text=\col] at (\c,1.49) {$g_{\lab}$};
\fi
}
\foreach \x/\lab in {0/0,0.333333/{L/3},0.666667/{2L/3},1/L}{\draw (\x,0.035) -- (\x,-0.035) node[below] {$\lab$};}
\end{scope}}
\end{tikzpicture}
\end{minipage}
}

\qrfThreeArcs{6}{L/6}\hfill
\qrfThreeArcs{30}{3L/10}

\caption{Three regional references on a circle, with periodic identification between $x=0$ and $x=L$. The pale blue, green and purple bands mark the supports of the averaging profiles for $A$, $B$ and $C$, respectively. Each yellow gap is the shared collar of two neighbouring regions. With the centres held fixed, the gap width decreases from $\ell=L/6$ (left) to $\ell=L/30$ (right), and the regional width increases from $w=L/6$ to $w=3L/10$. The normalised averaging profiles broaden and their peaks decrease (upper row). The charge plateaux widen over the same regions, and the transitions between them narrow (lower row). In each yellow gap the two neighbouring charge profiles sum to one, including across the periodic boundary.}
\label{fig:three-regional-profiles}

\endgroup
\end{figure}
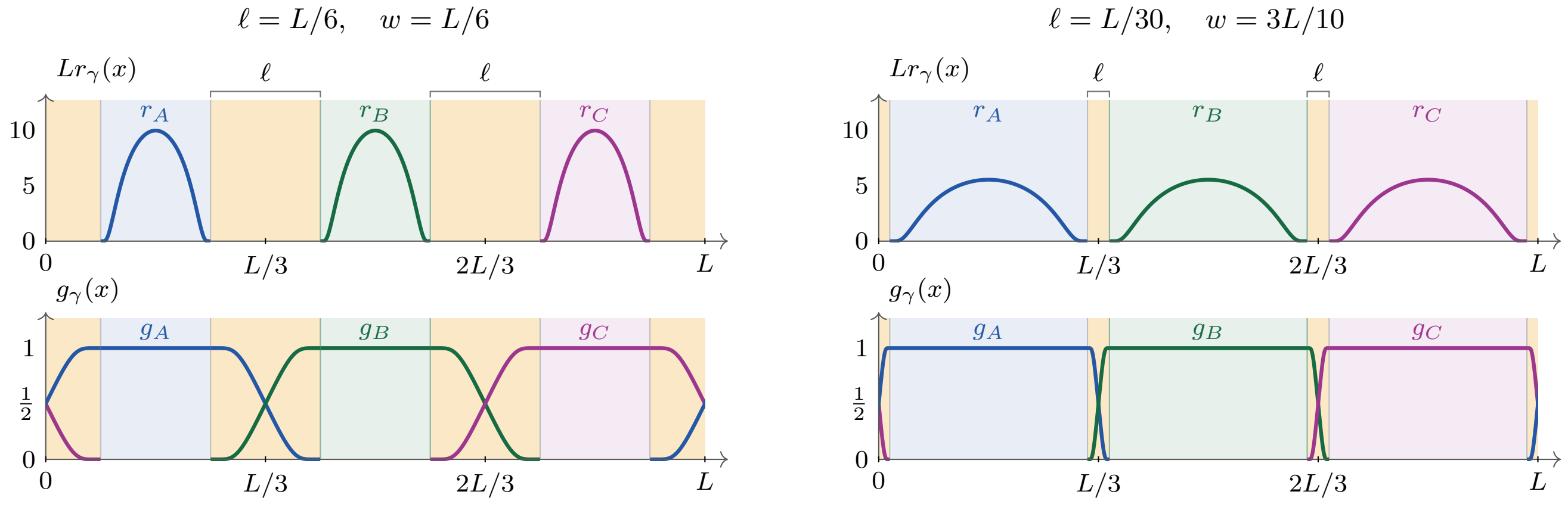

The collar width $\ell$ for the region $A$ depends on how the complement regions are placed -- and how much of the circle is not covered by any $r_\alpha$. A perfect decomposition into $A$ and the complement $\bar{A}$ does not exist precisely, but it can be achieved as a limit of increasing $w$ at fixed profile, reaching the zero-collar limit from above, $\ell \to 0^+$.We illustrate this limiting procedure for three regions in \Cref{fig:three-regional-profiles}.

The construction above identifies the collective field modes that serve as regional references. They are the field-theoretic counterparts of the regional collective pairs introduced in \cref{sec:lattice-construction}. 

\smallskip  \noindent\textbf{Ideal regional QRF.} The regional construction allows us to  define a regional QRF through a canonical collective mode $(\hat\Phi_\alpha,\hat\Pi_\alpha)$, as in the discrete case.  Its reference coordinate defines a PVM that is covariant under field shifts -- the regional modes are ideal QRFs.
\begin{equation}
    E_\alpha(\Delta)=\int_\Delta E_\alpha(\dd{\phi})=\int_\Delta \dd\Phi\,\ketbra{\Phi}{\Phi}_\alpha
    \qquad \Delta\subseteq\mathbb R\,, \label{eq:spectral_measure}
\end{equation}
where the kets are generalised eigenstates and the integral is understood spectrally. Under the shifts generated by $U_\alpha(a)=e^{-ia\hat\Pi_\alpha/\hbar}$, this PVM is covariant, namely 
\begin{equation}
    \hat U_\alpha(a)E_\alpha(\Delta)\hat U^\dagger_\alpha(a)
    =E_\alpha(\Delta+a)\,.
\end{equation}

Let us recall that the global field shift acts as \eqref{eq:global-field-shift} 
\begin{equation}
\mathcal{T}_a[\hat\Phi(x)]\coloneqq \hat U(a)^\dagger \hat\Phi(x) \hat U(a)= \hat\Phi(x) + a\,\mathds{1}\,,\qquad \hat U(a) = e^{-ia\hat\Pi_0/\hbar}\,.
\label{eq:shift_action}
\end{equation}
Let $E_A$ be the spectral measure of the regional reference coordinate $\hat\Phi_A$, introduced in \eqref{eq:spectral_measure}. We can therefore write the relativisation map using the same spectral decomposition as $\hat\Phi_A$, with the shift determined by each reference reading $\Phi$,
\begin{equation}
\mathcal R_A[\hat O]=\int_{\mathbb R} E_A(\mathrm d\Phi)\,\mathcal{T}_{-\Phi}[\hat O]\,.
    \label{eq:relativization_map}
\end{equation}
Let us now consider a real, smooth smearing $f$, which does not necessarily correspond to a region. We write $I_f = \int_0^L f(x)\,\mathrm dx$ and define $\hat\Phi^{(A)}(f) \coloneqq \mathcal R_A[\hat\Phi(f)]$. We obtain
\begin{equation}
\hat\Phi^{(A)}(f) = \int_{\mathbb R}E_A(\mathrm d\Phi)\,\bigl(\hat\Phi(f) - \Phi I_f\,\mathds{1}\bigr) = \hat\Phi(f) - I_f\hat\Phi(r_A)\,.
\end{equation}
The last equality follows from $E_A(\mathbb R) = \mathds{1}$ and $\int_{\mathbb R} \Phi\,E_A(\mathrm d\Phi) = \hat\Phi_A$. Using $  \hat\Phi(r_A)=\hat\Phi_A$, we can write the result as $\hat\Phi(f - I_f r_A)$. Since $r_A$ is normalised, the effective smearing has zero integral,
\begin{equation}
    \int_0^L \bigl[f(x) - I_f r_A(x)\bigr]\,\mathrm dx = 0\,.
\end{equation}
Thus, $\hat\Phi^{(A)}(f)$ is invariant under global field shifts. We now look for a Hilbert space decomposition for the field analogous to the discrete case of \cref{eq:lattice-hilbert-space-factorization-kinematical}. On the lattice, we separated the collective pair of each region from its internal degrees of freedom by taking differences of site coordinates within that region, as discussed in \cref{sec:lattice-construction}. A difference between two site coordinates is always invariant under a global shift. If both sites belong to $A$, their difference is also invariant when $A$ alone is shifted. If only one site belongs to $A$, their difference is not invariant.

Using the regional field pairs $(\hat\Phi_\alpha,\hat\Pi_\alpha)$, we can use the relativisation map  of the ideal reference  to make an analogous distinction. Consider the relativisation map in \cref{eq:relativization_map}. If $f$ is supported entirely within $A$, the relational observable $\hat\Phi^{(A)}(f)$ is invariant under every regional shift and plays the same role of an internal bulk coordinate on the lattice. If $f$ extends beyond $A$, the relational observable $\hat\Phi^{(A)}(f)$ is not invariant under field shifts generated by charge profiles $\hat\Pi_\beta$ with $\beta \neq A$, instead. 

 We want to construct relational observables that are invariant under any regional field shift, denoted as $\hat{\Phi}_\perp$ and $\hat{\Pi}_\perp$. In appendix \ref{app:canonical_complement_construction} we carry out this construction for arbitrary smearing functions.  Together, $\hat{\Phi}_\perp$ and $\hat{\Pi}_\perp$ form the canonical complement of the regional reference pairs, and read 
 \begin{equation}
\hat\Phi_\perp(f)=\hat\Phi(f)-\sum_\alpha a_\alpha(f)\hat\Phi_\alpha \,, \qquad a_\alpha(f)=\int_0^L \dd x\, f(x) g_\alpha(x)\,,
\label{eq:complementary-field-reading}
\end{equation}
\begin{equation}
    \hat\Pi_\perp(h)=\hat\Pi(h)-\sum_\alpha b_\alpha(h)\hat\Pi_\alpha\,,\qquad b_\alpha(h)=\int_0^L \dd x \, h(x)r_\alpha(x)\, ,
    \label{eq:complementary-momentum-reading}
\end{equation}
where the sum runs over all regions.
Thus, we obtain the kinematical decomposition
\begin{equation}
\mathcal H_{\mathrm{kin}}
\simeq
\left(\bigotimes_\alpha\mathcal H_\alpha\right)
\otimes\mathcal F_{\perp}\,,
\qquad
\mathcal H_\alpha\simeq L^2(\mathbb R)\,. \label{eq:canonical_decomposition_field}
\end{equation}
where each $\mathcal H_\alpha$ carries one regional reference pair and $\mathcal F_{\perp}$ carries their canonical complement. We can further resolve the factor $F_{\perp}$, by identifying the modes that strictly belong to the bulk of $A$ and its complement $\bar{A}$. Note, however, that there exists some smeared variables, with support neither in $A$ nor $\bar{A}$ -- those defined by smearing functions that have non-zero support in the collar region. With this choice of bulk mode factors for $A$ and $\bar A$, we denote the remaining factor by $\mathcal{F}_{\mathrm{rest}}$ and write: 
\begin{equation}
\mathcal F_{\perp}
\simeq
\mathcal H_A^b\otimes\mathcal H_{\bar A}^b
\otimes\mathcal F_{\mathrm{rest}}\,.
\end{equation}
This is a crucial difference compared to the lattice case, and has repercussions in the localisation of the algebra. We discuss this aspect in the next section.

\section{Regional algebras}\label{sec:edge_mode_qrf_approaches}

In this section, we examine what information remains accessible when the description is restricted to a spatial subregion, highlighting the role of regional charges and their connection with QRFs and edge modes in local gauge theories.  In particular, to make contact with edge modes in gauge theories, we restrict the theory to a fixed charge sector. As discussed in section \ref{sec:lattice-construction}, this implements a
strong symmetry reduction, treating the global shift as a gauge redundancy. Comparing the discrete lattice and continuum field constructions, we explain how spatial localisation affects access to the regional charge and to coherences between its different sectors.

\smallskip  \noindent\textbf{Regional algebras and canonical modes.}
By restriction to a spatial region $A$ we formally mean a restriction to the (von Neumann) algebra $\mathcal{A}_A$ generated by observables that are entirely contained in $A$, which we refer to as \emph{regional algebras}. On the lattice, these algebras are generated by the positions and momenta of the oscillators at the regional sites. The regional centre-of-mass pair $(\hat{\phi}_A, \hat{\pi}_A)$ of  \cref{eq:discrete-regional-cm}, as well as the internal bulk relational variables, are entirely supported within the corresponding region and make up all its degrees of freedom. The canonical decomposition induced by the regional QRFs of Sec. \ref{sec:lattice-construction}, that is \cref{eq:phys-edge-discrete},
\begin{equation}
\mathcal{H}_{\mathrm{kin}}\simeq
        \bigl(\mathcal{H}_{A}\otimes\mathcal{H}_A^b \bigr)\otimes\bigl(\mathcal{H}_{\bar A}\otimes\mathcal{H}_{\bar A}^b\bigr)
\end{equation}
is therefore the one induced by $\mathcal{A}_A$ and its commutant.

In the continuum, the regional algebras are generated by smoothly smeared fields and momenta \footnote{For a set of operators $\mathcal S$, its commutant $\mathcal S^\prime$ consists of all bounded operators commuting with every element of $\mathcal S$. Here, the double commutant gives the smallest von Neumann algebra containing the indicated Weyl operators. See e.g. Sec.6 of ref.\,\cite{FewsterRejzner2020}.}
\begin{equation}
\mathcal A_\gamma=\left\{\hat W(f,h)\vcentcolon f,h\in C_c^\infty(\gamma,\mathbb R)\right\}^{\prime\prime}\,,\qquad \gamma\in\{A,\bar A\}\,,
\label{eq:regional-algebra-qft-definition}
\end{equation}
where $\hat W(f,h)=e^{i[\hat\Phi(f)+\hat\Pi(h)]}$ are the Weyl operators and $C_c^\infty(\gamma,\mathbb R)$ denotes real-valued, infinitely differentiable functions whose support is a compact subset of the open region $\gamma$ \footnote{The smooth profiles $f,g$ specify the generators, not every operator belonging to the algebra: the latter also contains their algebraic combinations and strong operator limits, which might not be smooth.}. Note that the canonical factorisation induced by regional QRFs in ~\cref{eq:canonical_decomposition_field}, which here reads
\begin{equation}
\mathcal H_{\mathrm{kin}}\simeq\mathcal H_A\otimes\mathcal H_{\bar A}\otimes\mathcal F_{\perp}\,,
\label{eq:kin-hilb-space-decomposition-regional}
\end{equation}
is \emph{not} a factorisation induced by the regional algebras $\mathcal A_A$ and $\mathcal A_{\bar A}$. Indeed, the factor $\mathcal H_A$ carries the collective canonical pair $(\hat\Phi_A,\hat\Pi_A)$: while $\hat\Phi_A$ is supported within $A$, its conjugate momentum $\hat\Pi_A$ extends into the surrounding collar. As shown in Sec.~\ref{sec:regional_construction_conditions}, this extension is necessary to avoid UV divergences. In the discrete case, this problem does not arise: the observable algebra of a lattice region $A$ is a type-I factor, a von Neumann algebra isomorphic to $\mathcal B(\mathcal H)$, the algebra of all bounded operators on a Hilbert space $\mathcal H$. On the lattice, this identification is realised by $\mathcal A_A=\mathcal B(\mathcal H_A)\otimes\mathds 1_{\bar A}$. In continuum QFT, strictly local algebras are generally not type-I factors and therefore cannot be identified with the full operator algebra of a regional Hilbert-space factor~\cite{FewsterRejzner2020}. In appendix~\ref{app:regional_algebras_are_not_of_type_I}, we prove that the regional algebras defined in \cref{eq:regional-algebra-qft-definition} are not of type-I and relate the obstruction to the UV divergence of the sharp regional charge $\hat\Pi_A$, which requires smoothing through a collar.

\smallskip  \noindent\textbf{Regional charges and edge modes} We now examine how the global constraint ties together these regional canonical modes. Given a region A and its complement $\bar A$, we consider the constraint $\hat\Pi_0=\hat\Pi_A+\hat\Pi_{\bar A}=0$ \footnote{The same discussion applies to any fixed total-charge sector.}. This does not fix the regional charges individually, but requires them to compensate each other. The canonical factorization of \cref{eq:kin-hilb-space-decomposition-regional} leads to
\begin{equation}
    \mathcal H_{\mathrm{phys}}^{(0)}
    \simeq
    \int_{\mathbb R}^{\oplus} \dd \pi_A\,
    \left(
        \mathcal H_A^{(\pi_A)}
        \otimes
        \mathcal H_{\bar A}^{(-\pi_A)}
    \right)
    \otimes \mathcal F_{\perp}\,,
\label{eq:canonical_TPS_two_regions_field}
\end{equation}
The global constraint matches opposite regional charges, but does not remove coherences between different values of $\pi_A$. Detecting these coherences requires physical observables that do not commute with $\hat\Pi_A$. Since $\hat\Pi_0=\hat\Pi_A+\hat\Pi_{\bar A}$, global invariance then implies they do not commute with $\hat\Pi_{\bar A}$ either
\begin{equation}
[\hat O,\hat\Pi_0]=0,\quad [\hat O,\hat\Pi_A]\ne0 \quad\implies\quad [\hat O,\hat\Pi_{\bar A}]=-[\hat O,\hat\Pi_A]\ne0\,.
\end{equation}
Since, by construction, every observable supported entirely in $A$ commutes with $\hat\Pi_{\bar A}$\footnote{The charge profile $g_{\bar A}$ vanishes throughout $A$, and $g_A$ vanishes throughout $\bar A$.}, it follows that detecting coherences in the regional charge $\pi_A$ requires physical observables that are non-local.

This is even more evident in the factorisation that is adapted to the symmetry. One can isolate the relative degree of freedom between the two regions by defining
\begin{equation}
\begin{aligned}
\hat\Phi_+
&=\frac{\hat\Phi_A+\hat\Phi_{\bar A}}{2}\,,
&
\hat\Pi_0
&=\hat\Pi_A+\hat\Pi_{\bar A}\,,
\\
\hat\Phi_{\mathrm{edge}}
&=\hat\Phi_A-\hat\Phi_{\bar A}\,,
&
\hat\Pi_{\mathrm{edge}}
&=\frac{\hat\Pi_A-\hat\Pi_{\bar A}}{2}\,.
\end{aligned}
\end{equation}
The pairs $(\hat\Phi_+,\hat\Pi_0)$ and $(\hat\Phi_{\mathrm{edge}},\hat\Pi_{\mathrm{edge}})$ are canonical, and each operator in one pair commutes with both operators in the
other\footnote{Note that $\Phi_+$ is not $\Phi_0$, since, due to the collar, $A\cup \bar{A}$ is not the whole ring.}. This change of canonical variables therefore induces the factorisation
\begin{equation}
    \mathcal H_A \otimes \mathcal H_{\bar A}
    \simeq \mathcal H_+ \otimes \mathcal H_{\mathrm{edge}}\,.
\end{equation}
In the zero charge sector, the edge momentum satisfies $\hat\Pi_{\mathrm{edge}}=\hat\Pi_A=-\hat\Pi_{\bar A}$. Thus, imposing the constraint and hence removing the factor $\mathcal{H}_+$, the physical Hilbert space in \cref{eq:canonical_TPS_two_regions_field} decomposes as follows
\begin{equation}
\mathcal H_{\mathrm{phys}}
\simeq
\mathcal H_{\mathrm{edge}}
\otimes\mathcal F_{\perp}\,,
\end{equation}
The subscript “edge” is because of the analogy with boundary edge mode in local gauge theories \cite{DonnellyFreidel2016,CarrozzaHohn2022}: $\hat{\Pi}_{\mathrm{edge}} = \hat{\Pi}_A$, the generator of a relative shift between A and $\bar A$, is the analogue of a “corner symmetry generator” or "boundary charge" and $\hat{\Phi}_{\mathrm{edge}}$ is the conjugate “Goldstone-like” relational variable encoding how the regions are related. Physically, the collective mode $\hat{\Phi}_{\bar A}$ of the complement is used as a QRF for the collective mode $\hat{\Phi}_{A}$ of the sub-region A.

The same structure appears in the discrete construction of Sec. \ref{sec:lattice-construction}, where the “edge mode” corresponds to the canonical collective pair $(\hat\phi_A-\hat\phi_{\bar A},\hat\pi_A)$, consisting of the relative centre-of-mass position and the total momentum of the oscillators in the region $A$.

\smallskip  \noindent\textbf{Restriction to a subregion.} We can now restrict the description to region $A$, i.e. to the observables belonging to the regional algebra $\mathcal{A}_A$. By construction, they satisfy
\begin{equation}
    [\hat O_A,\hat\Pi_{\bar A}]=0\,,\implies [\hat O_A,\hat\Pi_0]=[\hat O_A,\hat\Pi_A]\,,  \forall\,\hat O_A\in\mathcal{A}_A\,.
\label{eq:local_observables_commutation_rel}
\end{equation}
Thus, every globally invariant observable supported in $A$ commutes with $\hat\Pi_A$ as well. Restricting to the invariant regional algebra
\begin{equation}
\mathcal{A}_A^{\mathrm{inv}} = \bigl\{\hat O_A\in\mathcal{A}_A \vcentcolon [\hat O_A,\hat\Pi_0]=0\bigr\}
\end{equation}
makes coherences between different regional charge sectors inaccessible, without fixing the charge to any particular value. This is the algebraic structure underlying \emph{weak symmetry reduction}. The same restriction on accessible information can be implemented by imposing shift symmetry ``weakly'', through an incoherent group average \cite{BartlettRudolphSpekkens2007, AugustinTotalCharge}. This illustrates how a strong global constraint can give rise to a weakly reduced regional description. In general, fixing the total charge does not fix the charge of a subsystem: the global constraint remains satisfied, while regional invariant observables cannot detect coherences between different regional charge sectors.

The difference between the discrete and continuum settings concerns access to the charge itself, i.e. whether $\hat \Pi_A$ strictly belongs to the regional algebra. In the continuum, $\hat\Pi_A$ extends into the collar and is therefore not entirely supported in $A$. Its value is inaccessible within $\mathcal A_A^{\mathrm{inv}}$, but becomes measurable if we enlarge this algebra by including the smooth regional charge. In the enlarged algebra, $\hat\Pi_A$ is the centre: its value can be measured, while coherences between its sectors remain inaccessible. The charge is thus retained as a classical label, without being fixed to a particular value. In the discrete setup, the total momentum $\hat\pi_A$ of the lattice region is entirely supported in $A$. It is therefore already an accessible central observable of the invariant regional algebra. Additionally, regional algebras are type I and provide a canonical factorisation of the kinematical Hilbert space into independent factors. Imposing the global constraint $\hat\pi_T=\hat\pi_A+\hat\pi_{\bar A}=0$ then gives the decomposition
\begin{equation}
\mathcal H_{\mathrm{phys}} \simeq \int_{\mathbb R}^{\oplus} \dd\pi_A\, \left(\mathcal H_A^{(\pi_A)} \otimes \mathcal H_{\bar A}^{(-\pi_A)} \right)\,.
\label{eq:hilbert-space-centre-of-mass-decomposition}
\end{equation}
The restriction to $\mathcal{A}^{\mathrm{inv}}_A$ corresponds to a partial trace over the complement factor, which gives physical states in the form
\begin{equation}
\hat \rho_A = \int \dd\pi_A\, \hat \rho_A^b(\pi_A)\otimes\ketbra{\pi_A}{\pi_A}_{A}\,,
\label{eq:discrete-regional-charge-register}
\end{equation}
where $\hat\rho_A^b(q)$ contain the bulk degrees of freedom. The regional charge is retained as a classical label, while the bulk degrees of freedom remain quantum within each sector. Again, the same description can be implemented by imposing translation symmetry ``weakly'' on region A.

This discussion shows that the regional charge can be seen either as the charge of a subregion within the larger constrained system, analogous to a ``corner symmetry'' generator in gauge theory, or as the total, physical charge if we take $A$ itself to be the system of interest. Taking the latter point of view connects this discussion to the role of the total charge ~\cite{Krummthirdparticle,CastroRuizGalleyLoveridge2025InterpretingQRF}. in different approaches to QRFs. Enlarging the physical observables to the collar in the continuum case amounts to considering the charge as a measurable quantity. In this case, retaining all charge sectors gives the Abelian extra-particle description \cite{CastroRuizOreshkov2025RelativeSubsystems}, while selecting a single sector (possibly also different to zero~\cite{deLaHametteKabelBrukner2025}) gives the fixed-charge description used in the perspective-neutral approach~\cite{VanrietveldeHohnGiacominiCastroRuiz2020}. Requiring the observables to be strictly localised within a region, instead, amounts to considering the charge as inaccessible. In this case, we recover the operational approach~\cite{CaretteGlowackiLoveridge2025, FewsterEtAl2025}, in which states that agree on all available measurements are identified. This occurs in the continuum when measurements are confined to the interior of $A$, and on the lattice if the observer has only access to bulk observables.  

\smallskip  \noindent\textbf{Extra-particle and edge modes.} Finally, the double role of the regional charge highlighted here suggests a connection between edge modes and the so-called ``extra particle'' in the theory of QRFs. Indeed, for abelian symmetries, the commutative algebra generated by the total charge corresponds to ``extra-particle'' degree of freedom \cite{CastroRuizOreshkov2025RelativeSubsystems,deLaHametteKabelBrukner2025,AugustinTotalCharge,SahdoCastroRuiz2026}, realising the connection explicitly in our model. Although the comparison with boundary charges in local gauge theories remains an analogy, our simple model provides some intuition: the charge records the matching condition $\hat\Pi_A+\hat\Pi_{\bar A}=0$ and must be retained to describe how the two regions can be combined. It would be interesting to investigate how this connection extends to non-Abelian symmetries, both on the lattice and in field theory, where the extra-particle algebra can contain quantum degrees of freedom beyond a classical charge label.

\section{Quantum reference frame transformations}\label{sec:quantum_reference_frame_trasformations}
For two complementary regions, the preceding section identified the relative amplitude $\hat\Phi_{\mathrm{edge}}=\hat\Phi_A-\hat\Phi_{\bar A}$ as the coordinate of the relational edge mode. This observable is inaccessible within either invariant regional algebra alone. It measures the shift in the origin of the field amplitude when we change the reference from $\bar A$ to $A$. We now consider this change of reference for any two regional QRFs and construct the unitary that implements it on the relative amplitudes and their conjugate momenta.

Take two reference modes $A,B$ and denote the remaining collective modes by $\alpha\in\Sigma$. Their charges satisfy
\begin{equation}
    \hat \Pi_0=\hat\Pi_A+\hat\Pi_B+\sum_{\alpha\in\Sigma}\hat\Pi_\alpha\,.
\end{equation}
Using $A$ as reference, define
\begin{equation}
    \hat\Phi_{B|A}=\hat\Phi_B-\hat\Phi_A\,,\qquad \hat\Phi_{\alpha|A}=\hat\Phi_\alpha-\hat\Phi_A\,, \quad\alpha\in\Sigma\,.
\end{equation}
The canonical variables relative to $A$ are
\begin{equation}
    (\hat\Phi_A,\hat \Pi_0)\,, \qquad (\hat\Phi_{B|A},\hat\Pi_B)\,, \qquad (\hat\Phi_{\alpha|A},\hat\Pi_\alpha)_{\alpha\in\Sigma}\,, \qquad  \mathcal O_\perp\,,
\end{equation}
where $\mathcal O_\perp$ denotes the complementary canonical observables. This pair reorganises the factors $\mathcal H_\alpha$ in \cref{eq:canonical_decomposition_field}. Recall that each such factor carries a single regional reference mode, whose conjugate charge extends into the collar, rather than the full algebra of observables strictly localised in $\alpha$.  The regional pairs $(\hat\Phi_\alpha,\hat\Pi_\alpha)$ are replaced by one global pair and a set of relative canonical pairs. Consequently, the same kinematical Hilbert space admits the factorisation
\begin{equation}
\mathcal H_{\mathrm{kin}}\simeq\mathcal H_{\mathrm{glob},}^{(A)}\otimes\mathcal H_B^{(A)}\otimes\left(\bigotimes_{\alpha\in\Sigma}\mathcal H_\alpha^{(A)}\right)\otimes\mathcal F_{\perp}\, \ ,
\end{equation}
Here $\mathcal H_{\mathrm{glob},}^{(A)}$ carries $(\hat\Phi_A,\hat \Pi_0)$, while $\mathcal H_B^{(A)}$ and $\mathcal H_\alpha^{(A)}$ carry the corresponding relative pairs. The superscript $(A)$ identifies the reference-adapted representation. The canonical complement is unchanged. The coordinate $\hat\Phi_A$ is well defined on the kinematical space and satisfies $[\hat\Phi_A,\hat\Pi_0]=i\hbar$. Therefore, it does not preserve a fixed-charge sector. Upon constraint reduction, this global coordinate is removed together with its conjugate charge. The relative coordinates commute with $\hat\Pi_0$ and remain physical observables.

Choosing $B$ as reference gives instead the chart
\begin{equation}
    (\hat\Phi_B,\hat\Pi_0)\,, \qquad (\hat\Phi_{A|B},\hat\Pi_A)\,, \qquad (\hat\Phi_{\alpha|B},\hat\Pi_\alpha)_{\alpha\in\Sigma}, \qquad \mathcal C_\perp\,,
\end{equation}
where
\begin{equation}
\hat\Phi_{A|B}=-\hat\Phi_{B|A}\,, \qquad \hat\Phi_{\alpha|B} =\hat\Phi_{\alpha|A}-\hat\Phi_{B|A}\,.
\end{equation}
The charge of region $A$ is reconstructed through
\begin{equation}
\hat\Pi_A=\hat\Pi_0-\hat\Pi_B-\sum_{\alpha\in\Sigma}\hat\Pi_\alpha\,.
\end{equation}

\begin{figure}[htbp]
\centering
\begingroup
\tikzset{
  site/.style={
    circle,draw=black,fill=black,inner sep=2pt
  },
  refsite/.style={
    circle,draw=black,fill=red!70,inner sep=3pt
  },
  regionbox/.style={
    draw=gray!55,rounded corners=3pt,line width=0.6pt
  },
  lab/.style={font=\small},
  vector/.style={->,thick,blue!70!black},
  edgevec/.style={->,thick,orange!85!black},
  svec/.style={->,thick,green!50!black}
}
\resizebox{0.75\linewidth}{!}{
\begin{tikzpicture}[scale=1.02,>=Latex]

\draw[regionbox] (-0.15,-0.15) rectangle (1.95,2.55);
\draw[regionbox] (4.25,-0.15) rectangle (6.35,2.55);
\node[lab,above=5pt] at (0.9,2.55) {$A$};
\node[lab,above=5pt] at (5.3,2.55) {$B$};

\draw[->,thin] (-0.35,0)--(9.65,0)
  node[right=5pt,lab] {$x$};
\draw[->,thin] (-0.65,-0.2)--(-0.65,3.15)
  node[above=5pt,lab] {$\Phi$};

\draw[thick,domain=0:9,samples=180]
  plot (\x,{
    1.4-0.6*cos(360*(\x-2)/9)
    +0.4*sin(720*(\x-2)/9)
  });

\node[lab,anchor=west, yshift=-0.1cm] at (2.15,0.5) {$\Phi(x)$};
\node[lab,below=5pt] at (0,0) {$0$};
\node[lab,below=5pt] at (9,0) {$L$};

\draw[gray!75,dashed]
  (-0.25,0.611536)--(9.45,0.611536);
\draw[gray!75,dashed]
  (-0.25,1.428483)--(9.45,1.428483);

\node[refsite] (field-XA) at (0.9,0.611536) {};
\node[refsite] (field-XB) at (5.3,1.428483) {};
\node[lab,red!70!black,anchor=west]
  at (9.5,0.611536) {$\Phi_A$};
\node[lab,red!70!black,anchor=west]
  at (9.5,1.428483) {$\Phi_B$};

\draw[gray!45,dotted] (7.8,0)--(7.8,2.131299);
\node[site] (field-f) at (7.8,2.131299) {};
\node[lab,above=5pt] at (field-f.north) {$\Phi(f)$};

\draw[edgevec] (3.7,0.611536)--(3.7,1.428483)
  node[midway,left=4pt,lab] {$\Phi_{B|A}$};
\draw[vector] (7.8,0.611536) to[bend right=18]
  node[pos=0.25,right=3pt,lab]
  {$\Phi^{(A)}(f)$} (field-f);
\draw[svec] (7.8,1.428483) to[bend left=25]
  node[midway,left=5pt,lab,xshift=0.1cm]
  {$\Phi^{(B)}(f)$} (field-f);

\draw[gray!45,thin] (-0.15,-1.05)--(9.2,-1.05);

\foreach \centre/\radius/\profile/\shade in
  {0.9/0.75/r_A/red!70!black,
   5.3/0.75/r_B/red!70!black,
   7.8/0.6/f/black}{
  \filldraw[
    draw=\shade,fill=\shade,fill opacity=0.10,thin
  ]
    ({\centre-\radius},-1.05)
    -- plot[
      domain=-0.999:0.999,
      samples=80,variable=\u
    ]
      ({\centre+\radius*\u},
       {-1.05+0.55*exp(1-1/(1-\u*\u))})
    -- ({\centre+\radius},-1.05)--cycle;
  \node[lab,below=5pt,text=\shade]
    at (\centre,-1.05) {$\profile$};
}

\end{tikzpicture}
}
\endgroup
\caption{Continuum counterpart of
\cref{fig:regional-qrf-discrete}.
Heights represent illustrative expectation values, with hats and expectation brackets suppressed. Normalised profiles $r_A,r_B,f$ define the regional averages and the smeared field observable. Changing reference gives $\Phi^{(B)}(f)=\Phi^{(A)}(f)-\Phi_{B|A}$ at fixed $x$. The profiles below the graph use an illustrative vertical scale.}
\label{fig:regional-qrf-continuum}
\end{figure}
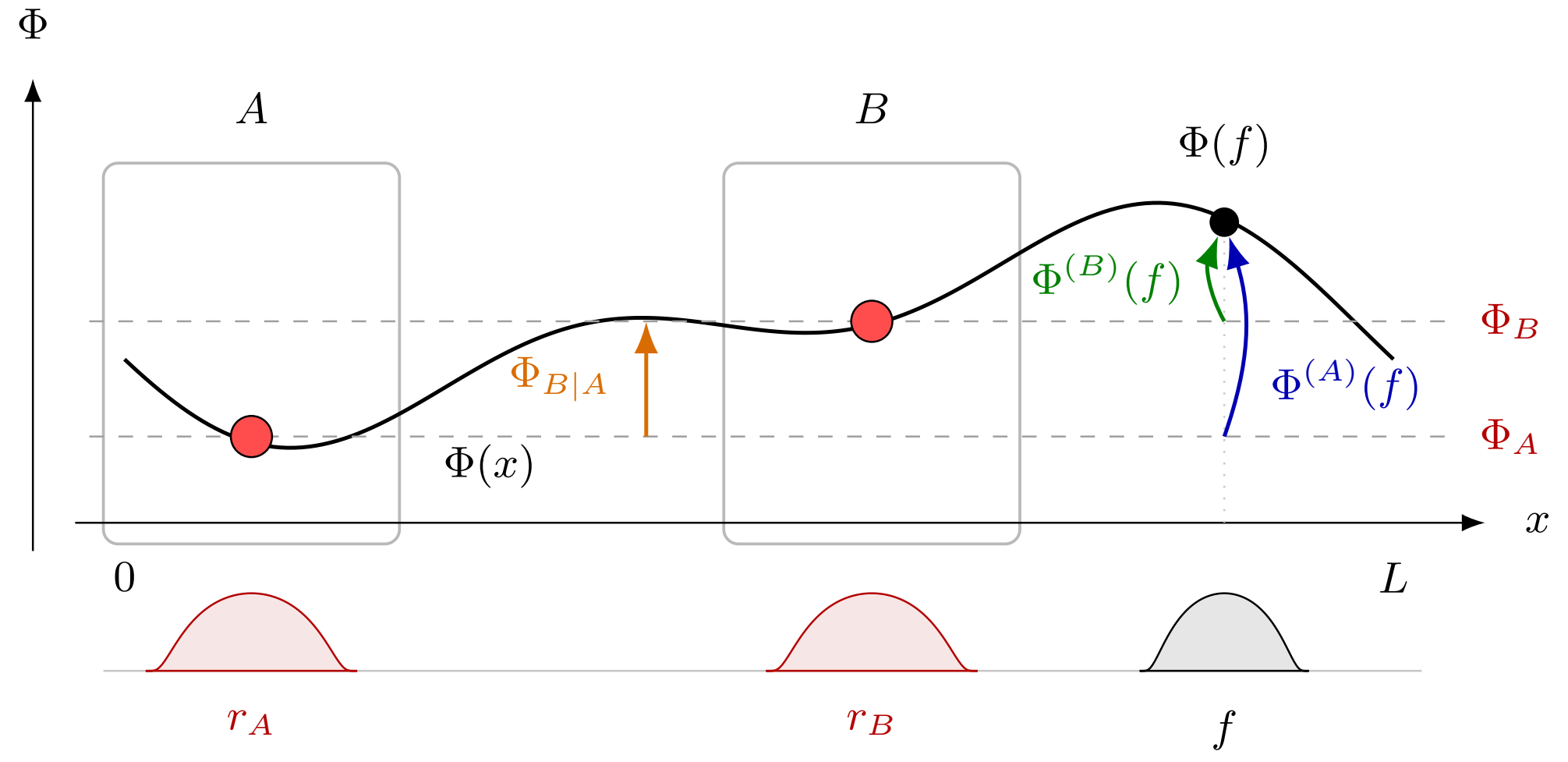

The transformation mapping from the perspective of $A$ to the perspective of $B$ is
\begin{equation}
\hat S_{A\to B}=\mathcal P_{A\to B}\hat U_{AB}\,,\qquad\hat U_{AB}=\exp\!\left[\frac{i}{\hbar}\hat\Phi_{B|A}\hat G_{AB}\right]\,,
\end{equation}
with
\begin{equation}
\hat G_{AB}=\sum_{\alpha\in\Sigma}\hat\Pi_\alpha-\hat\Pi_0=-(\hat\Pi_A+\hat\Pi_B)\,,
\end{equation}
and $\mathcal{P}_{A\rightarrow B}$ being the parity-swap operator~\cite{GiacominiCastroRuizBrukner2019}. Notice that this transformation is unitary and the presence of the global charge in the transformation is a fundamental ingredient for its unitarity.  The relational variables and the transformation are illustrated in \cref{fig:regional-qrf-continuum}.  The global charge label is carried unchanged, and the operator acts as the identity on $\mathcal F_{\perp}$. After the transformation, the relational observables are re-expressed as
\begin{equation}
\left\{
\begin{aligned}
    \hat\Phi_{\alpha|A}
    &\mapsto
    \hat\Phi_{\alpha|B}-\hat\Phi_{A|B}\,,
    &&\alpha\in\Sigma\,,\\
    \hat\Phi_{B|A}
    &\mapsto-\hat\Phi_{A|B}\,,\\
    \hat\Pi_\alpha^{(A)}
    &\mapsto\hat\Pi_\alpha^{(B)}\,,
    &&\alpha\in\Sigma\,,\\
    \hat\Pi_B^{(A)}
    &\mapsto
    \Pi_0-\hat\Pi_A^{(B)}
    -\sum_{\alpha\in\Sigma}\hat\Pi_\alpha^{(B)}\,.
\end{aligned}
\right.
\end{equation}
The superscripts on the momenta indicate their reference representations. The complement transforms trivially. The action on arbitrary relative field observables is
\begin{equation}
\hat\phi^{(A)}(f)\mapsto \hat\phi^{(B)}(f)-I_f\hat\Phi_{A|B}\,,\qquad I_f=\int_0^Ldy\,f(y)\,.
\end{equation}

We now return to $\Pi_0=0$. Constraint reduction removes the global pair and leaves
\begin{equation}
    \mathcal H_{\mathrm{phys}}^{(0)}\simeq \mathcal H_B^{(A)} \otimes \left( \bigotimes_{\alpha\in\Sigma}\mathcal H_\alpha^{(A)}\right) \otimes\mathcal F_{\perp}\,.
\end{equation}
For two complementary regions $A,\bar A$, the controlled unitary is
\begin{equation}
    \hat U_{A\bar A}=\exp\!\left[-\frac{i}{\hbar}\hat\Phi_{\bar A|A}\Pi_0\right]\,.
\end{equation}
In a fixed-charge sector $\Pi_0=Q$, the transformation becomes
\begin{equation}
    \hat S_{A\to\bar A}^{(Q)}=\mathcal P_{A\to\bar A}\exp\!\left[-\frac{iQ}{\hbar}\hat\Phi_{\bar A|A}\right]\,.
\end{equation}
At $Q=0$, it reduces to a parity-swap on the surviving relative collective mode,
\begin{equation}
    \hat S_{A\to\bar A}^{(0)}=\mathcal P_{A\to\bar A}\,,
\end{equation}
with the identity on the canonical complement.

These transformations are the natural generalisation of the reference frame transformations for a single-particle reference frame~ by~\cite{GiacominiCastroRuizBrukner2019, DeLaHametteGalley2020, VanrietveldeHohnGiacominiCastroRuiz2020, DeLaHametteEtAl2021, CastroRuizOreshkov2025RelativeSubsystems, CastroRuizGalleyLoveridge2025InterpretingQRF}. This is more evident for the lattice, where the regional transformations have the same canonical form, with $(\hat\Phi_\alpha,\hat\Pi_\alpha,\Pi_0)$ replaced by $(\hat\phi_\alpha,\hat\pi_\alpha,\pi_T)$ and the field complement replaced by the discrete bulk degrees of freedom.

\section{The energy cost of precision}\label{sec:energy_cost_of_precision}

We now turn the attention to a simple metrology task in field theory, in order to elucidate the power of having access to a quantum reference frame when estimating a relational observable.  This bridges our discussion to experiments involving condensed-matter systems and, in particular, matter-wave interferometry with Bose–Einstein condensates. 

Let us see why the massless field is relevant in this context. Collective motion in a Bose-Einstein condensate is encoded in the effective global wave-function
\begin{equation}
    \psi(x)=\sqrt{n(x)} e^{i\theta(x)} \, ,
\end{equation}
where $n(x)$ represents the number density of atoms in the condensate, and $\theta(x)$ is a real phase. In the limit of small phase fluctuations $\Phi(x)$ around a fixed classical profile $\theta_0$ (i.e. $\theta-\theta_0= \Phi$), the effective dynamics of the fluctuations and number density decouple, with the former being described by a massless scalar field on the circle~\cite{Eckel:2020qee}. 

\begin{figure}[htbp]
\centering
\definecolor{phaseInk}{HTML}{273238}
\definecolor{phaseGrey}{HTML}{53626B}
\definecolor{phaseAxis}{HTML}{7B898F}
\definecolor{phaseRing}{HTML}{DFE5E7}
\definecolor{phaseGreen}{HTML}{236E3B}
\definecolor{phaseGreenFill}{HTML}{A8DAA5}
\definecolor{phaseViolet}{HTML}{494AA0}
\definecolor{phaseVioletFill}{HTML}{C6C1EA}
\definecolor{phaseBeam}{HTML}{FFF1D9}
\definecolor{phaseBeamCore}{HTML}{FAE0B8}
\definecolor{phaseLens}{HTML}{BD762F}

\resizebox{\linewidth}{!}{\begin{tikzpicture}[
    x=1bp,y=1bp,>=stealth,
    line cap=round,line join=round,
    inner sep=0pt,outer sep=0pt,text=phaseInk,
    axis/.style={draw=phaseAxis,line width=.75bp,->},
    background/.style={draw=phaseGrey,line width=1bp,line cap=butt},
    target/.style={draw=phaseGreen,line width=2bp,line cap=butt},
    reference/.style={draw=phaseViolet,line width=2bp,line cap=butt}
]
\path[use as bounding box] (0,0) rectangle (590,233);
\newcommand{\phaseBaseline}[2]{
    \pgfmathtruncatemacro{\phaseGaps}{max(1,round((#2-#1-3)/5.8))}
    \pgfmathsetlengthmacro{\phaseGap}{((#2-#1-3)/\phaseGaps-3)*1bp}
    \draw[background,dash pattern=on 3bp off \phaseGap]
        (#1,169) -- (#2,169);
}
\foreach \offset/\stage in {0/0,304/1}{
    \begin{scope}[shift={(\offset,0)}]
        \draw[axis] (27,151) -- (266,151) node[right=2bp] {$x$};
        \draw[axis] (27,151) -- (27,220) node[left=4bp] {$\Phi$};
\foreach \pos/\lab in {27/0,258/L}{
            \draw[phaseAxis] (\pos,153) -- (\pos,149)
                node[below=2bp,text=phaseGrey] {$\lab$};
        }
\ifnum\stage=0\relax
            \phaseBaseline{28}{163.29}
\else \phaseBaseline{28}{35.86}
\foreach \side in {-1,1}{
\begin{scope}[shift={(74.36,169)},xscale=\side]
                    \draw[target] (0,38)
                        .. controls (1.92,38) and (3.84,35.11) .. (5.77,30.77)
                        .. controls (7.69,26.43) and (9.62,20.83) .. (11.54,16.01)
                        .. controls (14.11,9.58) and (16.68,4.81) .. (19.25,2.54)
                        .. controls (21.81,.27) and (24.38,.10) .. (26.95,.03)
                        .. controls (30.80,0) and (34.65,0) .. (38.50,0);
\end{scope}
            }
\phaseBaseline{112.86}{163.29}
\draw[phaseAxis,line width=.48bp]
(74.36,207) -- (130.64,207);
\draw[<->,line width=.85bp]
(128.64,169) -- (128.64,207)
node[midway,right=6bp] {$\Phi_{f\mid A,w}$};
\fi
\draw[reference] (163.29,169) -- (197.94,169);
\phaseBaseline{197.94}{258}
\begin{scope}[shift={(121,67)}]
\ifnum\stage=0\relax
\fill[phaseBeam]
(98,53) -- (101,43) -- (-13.8:43.9bp)
arc[start angle=-13.8,end angle=46.2,radius=43.9bp]
-- cycle;
\fill[phaseBeamCore]
(98,51) -- (100,45) -- (-.3:43.9bp)
arc[start angle=-.3,end angle=32.7,radius=43.9bp]
-- cycle;
\begin{scope}[shift={(99,48)},rotate=28]
\draw[phaseGrey,fill=phaseRing,line width=.75bp]
(1,-3.7) -- (3,-5) -- (19,-5) -- (21,-3.4)
-- (21,3.4) -- (19,5) -- (3,5) -- (1,3.7)
-- cycle;
\foreach \s in {15,17.5}
\draw[phaseAxis] (\s,-4.5) -- (\s,4.5);
\draw[phaseGrey,fill=phaseAxis]
(0,-3.3) rectangle (3,3.3);
\draw[phaseLens,line width=1.1bp]
(0,-2.2) -- (0,2.2);
\end{scope}
\fi
\draw[phaseRing,line width=8.7bp]
(0,0) circle[radius=42bp];
\draw[phaseVioletFill,line width=8.7bp]
(-176.4:42bp)
arc[start angle=-176.4,end angle=-122.4,radius=42bp];
\ifnum\stage=1\relax
\draw[phaseGreenFill,line width=8.7bp]
(-13.8:42bp)
arc[start angle=-13.8,end angle=46.2,radius=42bp];
\fi
\draw[phaseGrey,line width=.8bp]
(0,0) circle[radius=42bp];
\draw[phaseViolet,line width=1.35bp]
(-176.4:42bp)
arc[start angle=-176.4,end angle=-122.4,radius=42bp];
\node[phaseViolet,anchor=east] at (-50,-28) {\large $r_A$};
\ifnum\stage=1\relax
\draw[phaseGreen,line width=1.35bp]
(-13.8:42bp)
arc[start angle=-13.8,end angle=46.2,radius=42bp];
\node[phaseGreen,anchor=west] at (52,-5) {\large $f$};
\fi
\end{scope}
\end{scope}
}
\draw[->,line width=1bp] (251,70) -- (308,70)
    node[midway,above=5bp] {$\tau$};
\end{tikzpicture}}

\caption{Relative phase before (left) and after (right) a short pulse of duration $\tau$ in a ring condensate. On the left, we depict the region $r_A$ that we use as our reference. The upper panel shows that the initial phase is constant. On the right, the laser imprints a local phase shift $\delta\theta(x)$ on the illuminated arc (green). We use the normalised laser intensity profile as the weight $f$ with which we average the phase over that arc. The peak in the upper right panel gives the target reading $\Phi(f)$, and the vertical arrow shows its difference from the reference reading $\Phi(r_{A,w})$ on the unilluminated arc (violet).}
\label{fig:ring-phase-before-after}
 \end{figure}
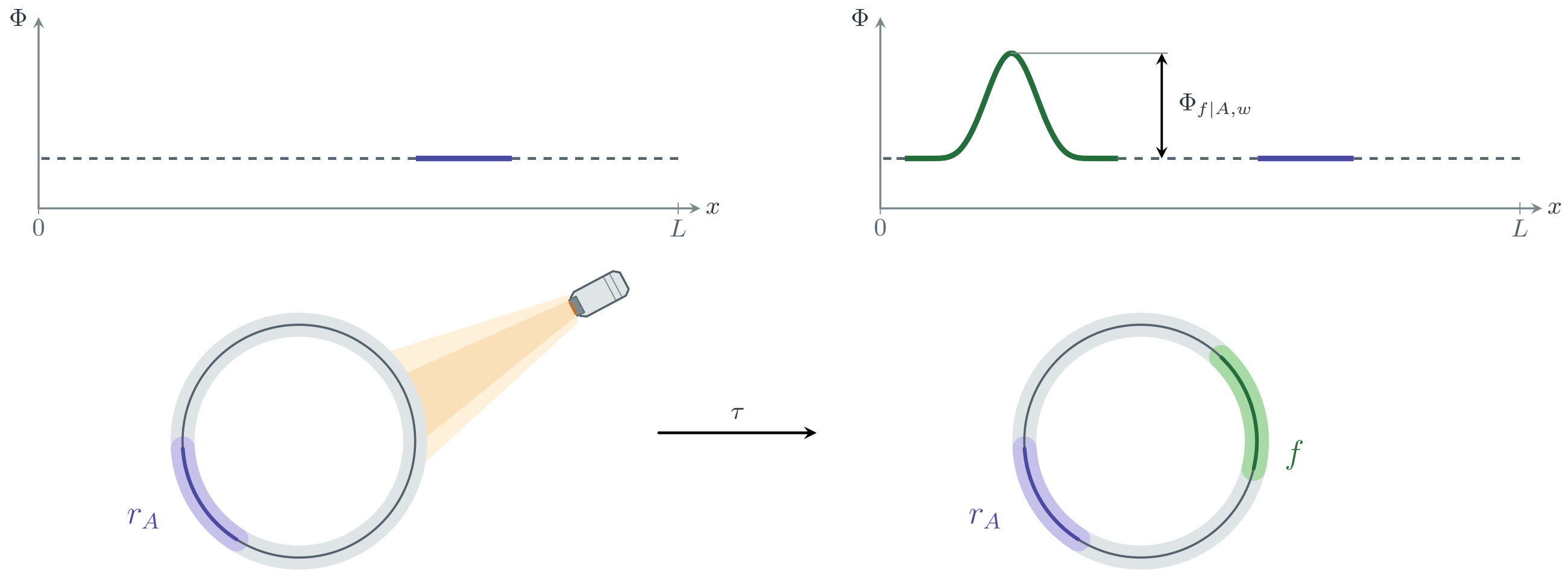

It is possible to dynamically imprint a phase-difference between different regions of the condensate. To achieve this, one introduces an external field that couples to the atomic density $|\psi|^2$ through a potential energy $U(x,t)$
\begin{equation}
    H_{\mathrm{ext}}=\int \dd[3]{x} \, U (x, t)|\psi  (x, t)|^2 \, .
\end{equation}
The time evolution with this Hamiltonian induces a phase $\Phi(x) \sim U\tau /\hbar$, in the small-coupling limit and short interrogation times $\tau$, during which the density $n(x)$ is taken to be constant. Simple examples of such an external field are the gravitational field (that couples to the mass density $m|\psi|^2$), or an off-resonant laser. If $U(x)$ is not constant on the circle (for example, it is a laser whose intensity profile induces some smooth regional smearing function $f$), a phase difference is induced~\cite{Denschlag2000}, as shown in~\cref{fig:ring-phase-before-after}. Estimating the energy difference between different regions of the condensate, encoded in the phase difference between the corresponding regions, is a simple metrology task. Indeed, such phase difference corresponds to the relational amplitude
\begin{equation}
    \hat \Phi_{f|A,w}\equiv\hat\Phi(f)-\hat{\Phi}_{A,w}
    =\hat\Phi(h_w)\,,\qquad h_w\equiv f-r_{A,w}\,,
\end{equation}
between the region $f$ shone by, for example, the laser and some reference region $A$ identified by the regional smearing $r_{A,w}$, where $w$ is the width. This phase readout is illustrated in \cref{fig:ring-phase-before-after}.

 We can therefore access the phase-shift by using part of the unaffected condensate as a reference frame~\cite{Simsarian2000}. Estimating $U$ precisely requires suppressing the quantum fluctuations on the $\hat \Phi_{f|A,w}$ operator. To see how such fluctuations affect the relational reading, it is useful to imagine $h_w(x)$ as a filter function for the observable of interest (here $f$ is normalised as well as $r_A$). Expanding the field in the Fourier basis functions $e_j(x)$ introduced in \cref{eq:real-fourier-mode-expansion}
\begin{equation}
\hat \Phi_{f|A,w}=\sum_j d_j(w)\hat q_j\,,\qquad d_j(w)=\int_0^L \dd x\,h_w(x)e_j(x) \, , \label{eq:expansion-application}
\end{equation}
with $d_j(w)$ the real Fourier coefficient of the filter function $h_w$, and $\hat q_j\propto \hat{a}_j+\hat{a}_j^\dagger$ the $j$-th mode amplitude. The vacuum fluctuations give a reference-dependent uncertainty on the observable of interest
\begin{equation}
\sigma^2_{\mathrm{vac}} \equiv\langle0|\hat \Phi_{f|A,w}^2|0\rangle =\frac{\hbar}{2}\sum_j\frac{d_j(w)^2}{\omega_j} \, ,
\end{equation}
where, recall $\omega_j = 2\pi vj/L$ are the discrete frequencies of the $j$-th mode.
This variance quantifies the unsharpness of the relative reading in the vacuum. A suitable preparation of the field, including the reference degrees of freedom and their correlations with the target, can reduce it, allowing a sharper estimation of $U$. This is a standard optimisation problem in metrology~\cite{Maccone2020squeezingmetrology}. 

Such a preparation requires energy. It is therefore natural to assess the advantage of a given preparation by asking how small the variance of the relative field amplitude can be made for fixed target and reference profiles and a given mean excitation-energy budget $E$. We measure $E$ relative to the lowest energy in the chosen total-charge sector $Q$. Since the uniform mode contributes the same energy $Q^2/(2L)$ to every state in that sector, $E$ is the energy available to excite the non-zero modes. The minimum variance is (see Appendix \ref{app:optimal_state} for details)
\begin{equation}
\sigma^2_{\mathrm{min}}(w;E) \equiv \inf_{\rho:\,\mathrm{Tr}(\rho\hat H_{\mathrm{exc}})\le E} \mathrm{Tr}\left[\rho \ \hat \Phi_{f|A,w}^2\right] \, ,
\end{equation}
where we have assumed $\mathrm{Tr}(\rho \hat \Phi_{f|A,w})=0$ since shifting the mean alone of $\Phi$ has no effects on the absolute size of the fluctuations. Introducing the energy constraint through a Lagrange multiplier $\mu\ge0$, we can therefore optimise the state of the field via the functional
\begin{equation}
\mathcal L_\mu(\rho) =\frac{1}{2}\operatorname{Tr}\!\left[\rho\hat \Phi_{f|A,w}^2\right] +\mu\left[\operatorname{Tr}(\rho\hat H_{\mathrm{exc}})-E\right]\,. \label{eq:lagrange_multiplaier}
\end{equation}
For $\mu>0$, division by $\mu$ identifies an auxiliary Hamiltonian,
\begin{equation}
\hat H_\lambda =\hat H_{\mathrm{exc}}+\frac{\lambda}{2}\hat \Phi_{f|A,w}^2=\frac{1}{2}\hat p^{T}\hat p+\frac{1}{2}\hat q^{T}K_\lambda\hat q-E_{\mathrm{vac}} \, ,
\end{equation}
where
\begin{equation}
K_\lambda=\Omega^2+\lambda dd^{T}\,, \qquad \Omega=\mathrm{diag}(\omega_j) \, ,
\end{equation}
and $\lambda=1/\mu$, and $E_{\mathrm{vac}}=\hbar/2\mathrm{Tr}(\Omega)$. Here and below, matrix expressions are strictly valid at a finite mode cutoff (see  appendix~\ref{app:optimal_state}). The problem then reduces to simply finding the ground state of the quadratic auxiliary Hamiltonian $\hat H_\lambda$, with $\lambda$ chosen so that its physical excitation energy equals $E$. Since $K_\lambda$ is positive definite, this ground state is unique.

Note that the optimisation scans over states that might require control of the quantum system on the whole ring, to be prepared from the vacuum. Indeed, as we shall see, the real advantage comes from allowing strong correlations between the reference and target regions, rather than optimising the state of the reference frame with local operations alone. A clear analogy of why this might be the case comes from historical reference frames literature, in which two-mode squeezing, via entanglement between the phase-space degrees of freedom of a reference frame and particle of interest, was leveraged to remove the effects of the quantum fluctuations on the trajectory of the particle~\cite{Polzik:2014xza, Moller:2017yzv}. Recently, sharpening of particle trajectories using QRFs was studied in Ref.\,\cite{Fadel:2026zup}.

\subsection{The advantage of squeezing}
Gaussian states are completely specified by their first moments and covariance matrix. At zero excitation energy, the only admissible state in the non-zero-mode sector is the physical vacuum, whose wave-functional, expressed in the real mode amplitudes $q_j$, is
\begin{equation}
    \Psi_0(q) =\mathcal{N} \exp\!\left[-\frac{1}{2\hbar}q^T\Omega q\right]\,, \qquad \expval{\hat q\hat q^T}_0=\frac{\hbar}{2}\Omega^{-1}\,, \qquad \expval{\hat p\hat p^T}_0=\frac{\hbar}{2}\Omega \, ,
\end{equation}
for some formally divergent normalisation constant $\mathcal{N}$. 

Adding energy can modify either the mean field or its fluctuations. A multimode coherent state has mean field and momentum centred on a chosen, classical, initial data $(\Phi_{\mathrm{cl}},\Pi_{\mathrm{cl}})$. However, its connected two-point functions remain those of the vacuum, providing no advantage. For optimality, the available energy must instead modify the two-point function in a way that suppresses the modes with the largest Fourier coefficients $d_j$ instead. As shown in Appendix~\ref{app:optimal_state}, the optimal preparation turns out to be the centred multimode squeezed vacuum
\begin{equation}
    \Psi_\lambda(q) =\mathcal{N}^\prime \exp\!\left[-\frac{1}{2\hbar}q^TK_\lambda^{1/2}q\right]\,,
\end{equation}
with covariances
\begin{equation}
    \Gamma_{qq}^{(\lambda)}=\frac{\hbar}{2}K_\lambda^{-1/2}\,, \qquad \Gamma_{pp}^{(\lambda)}=\frac{\hbar}{2}K_\lambda^{1/2}\,, \qquad \Gamma_{qp}^{(\lambda)}=0 \, .
\end{equation}
Equivalently, the field two-point function in space is
\begin{equation}
    \expval{\hat\Phi_{\mathrm{osc}}(x)\hat\Phi_{\mathrm{osc}}(y)}_\lambda =\frac{\hbar}{2}\sum_{j,k}e_j(x) \left[K_\lambda^{-1/2}\right]_{jk}e_k(y)\,.
\end{equation}
Choosing the unique $\lambda_E>0$ satisfying the energy constraint gives
\begin{equation}
    \sigma^2_{\mathrm{min}}(w;E) =\frac{\hbar}{2}d(w)^TK_{\lambda_E}^{-1/2}d(w)\,, \qquad E_w(\lambda_E)=E\,.
\end{equation}

The solution of the optimisation problem is intuitive. As a consistency check, at $E=0$ we have $\lambda_E=0$. Therefore, $\sigma_{\mathrm{min}}(w;0)=\sigma_{\mathrm{vac}}$, and we indeed recover the vacuum state result. Since (as we show in Appendix~\ref{app:optimal_state}) as $E$ increases, so does $\lambda$, given an unbounded energy budget, to minimise the variance the quantity $K_\infty \sim \lambda d d^T$ should be made large. Recall that the vector $d(w)$ encodes the Fourier coefficients of the effective filter smearing $h_w=f-r_w$. Hence, enhancing $K_\lambda$ in that subspace simply means suppressing the amplitudes of those modes in real space, which are controlled by $K$-inverse instead. The price to pay is, of course, that the orthogonal modes become highly enhanced. In particular, when $f$ is understood as another regional reference frame, it is the conjugate “edge” momentum $(\Pi_B-\Pi_A)/2$ that loses sharpness. This realises the phase-space interpretation of squeezing, where momentum uncertainty is sacrificed to sharpen spatial localisation.

Some qualitative features of the functional dependence of $\sigma_{\mathrm{min}}$ on $E$ and $w$ have a straightforward physical interpretation. As we show in Appendix~\ref{app:optimal_state}, explicit differentiation shows that additional energy always implies an enhancement in precision
\begin{equation}
    \frac{\partial \sigma^2_{\mathrm{min}}}{\partial E}= -\frac{2}{\lambda} <0 \, .
\end{equation}
Secondly, for small $w$, the leading contributions will simply be the logarithmic divergent vacuum fluctuations of the reference observable
\begin{equation}
    \sigma^2_{\mathrm{min}}(E,w) \propto \sum_{j=1}^{J^w} \frac{1}{\omega_j}\,,
\end{equation}
with the sum truncated at $J_w$, i.e. when $\omega_j$ reaches the length scale $v/w$ associated with the size of the reference.  Modes that oscillate many times within the arc contribute little to the average because their positive and negative contributions largely cancel. A narrower reference region therefore remains sensitive to higher-frequency modes, giving a larger $J_w$. The smooth averaging profile suppresses these modes gradually, so
$J_w$ represents an effective cutoff. 

\begin{figure}[htbp]
    \centering
    \begin{subfigure}[b]{0.48\textwidth}
        \centering
        \includegraphics[width=\linewidth]{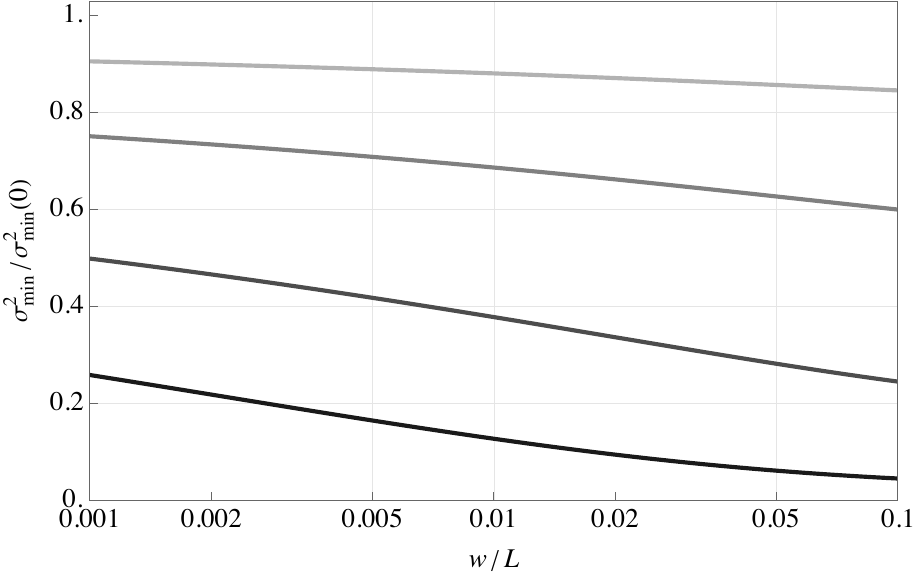}
        \caption{}
        \label{fig:advantage}
    \end{subfigure}
    \hfill
    \begin{subfigure}[b]{0.48\textwidth}
        \centering
        \includegraphics[width=\linewidth]{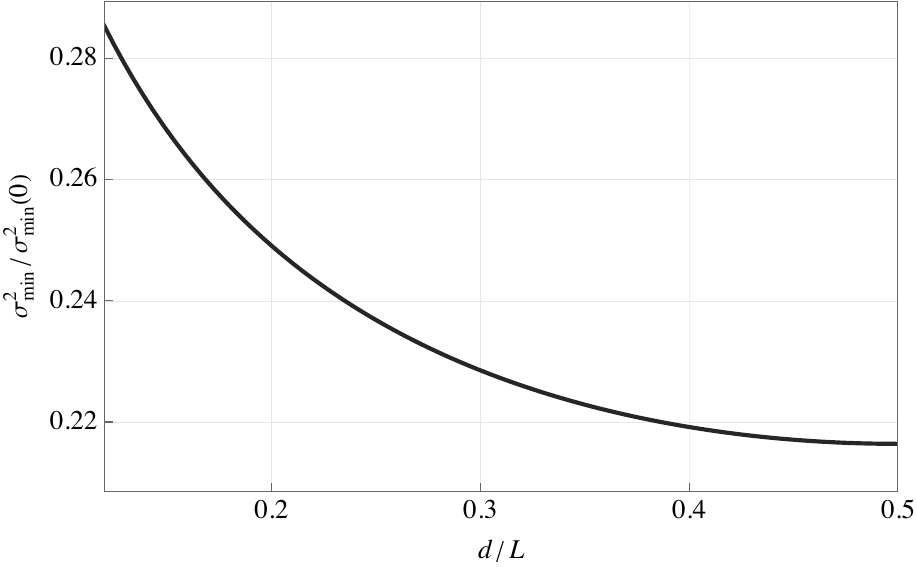}
        \caption{}
    \end{subfigure}

    \caption{(a) Fractional suppression of the variance of the relational amplitude
below its vacuum value, plotted against regional width on a logarithmic axis. Each curve fixes the dimensionless excitation-energy budget $EL/(\hbar v)=0.1,1,10,100$, from lighter to darker. Increasing the energy budget monotonically improves the precision at fixed width and, for the chosen family, increasing $w$ always yields an improved estimate. (b) The advantage of moving the reference region further away from the target observable, at fixed $EL/(\hbar v)=10$ and $w/L=0.12$.} 
    \label{fig:squeezing}
\end{figure}

There is an initial advantage in increasing the size of the region, since more and more modes contribute into the average that defines the reference observable, reducing its fluctuations. The precise functional dependence on $w$, however, depends on the choice of smearing functions used to identify the reference and target observables, and their relative spatial location. A numerical investigation of the advantage of preparing a squeezed state, compared to the vacuum, can be found in Figure~\ref{fig:squeezing} for the smearing function defined in ~\cref{sec:regional_construction_conditions}. In the right panel, we show numerically that Placing the reference region farther from the target region increases the fractional suppression of the variance relative to its vacuum value. This also has an intuitive physical explanation, as the separation $d$ between the centres of the target and regional references increases, the relational observable becomes sensitive to longer frequencies. For the same energy budget, squeezing those frequencies is relatively easier, providing an advantage.

Squeezed states are generically optimal in metrology problems~\cite{Maccone2020squeezingmetrology}. Our result is a natural extension of well-known protocols in quantum optics, where correlations between probe and reference modes, or among spatially separated sensing modes, reduce the noise of joint quadrature measurements \cite{Steinlechner:2012mjw,Zhuang:2018wqs}. The optimal squeezed state should be understood as the counterpart, in the continuum, of preparing a multidimensional entangled state between reference and target system. In this sense, our computation highlights the advantage of having access to quantum reference frames that admit enhanced quantum correlations with the system of interest.

\subsection{Two-party trade-off}
We now ask whether the optimal state for one choice of reference frame is also optimal for other frames. The answer is, generically, no. As we see now, there exists a trade-off at fixed energy between how well a relational observable of interest can be sharpened for two distinct regions $A$ and $B$ at the same time. To gain an intuition, define $\sigma^2_{w,A}$ and $\sigma^2_{w,B}$ to be the variance of the observable relationalised with respect to $A$ and $B$ respectively, and
\begin{equation}
     \sigma^2_{w,AB}=\mathrm{Var}\left(\hat{\Phi}(f)- \frac{\hat{\Phi}_A+\hat{\Phi}_B}{2}\right) \equiv \mathrm{Var}\left(\hat{\Phi}_{f|AB}\right) \,,
\end{equation}
the relative variance obtained with respect to an equally-weighted reference frame made up of the \emph{joint} region $A\cup B$. Realising $\Phi_{f|A,B}=\Phi_{f|AB} \pm \Phi_{A|B}/2$, and using standard properties of the variance, we see that
\begin{equation}
    \sigma^2_{w,A}-\sigma^2_{w,B} = 2 \ \mathrm{Cov}\left(\hat{\Phi}_{f|AB},\hat{\Phi}_{B|A} \right) \, .
\end{equation}
If $A$ and $B$ are equally good frames for the chosen observable, reaching the optimal value for both at fixed energy would imply the right-hand side vanishes. However, that is not generically the case. Optimising the field state for both relational observables at the same time means spending some of the energy budget to suppress the covariance between the average relational observable and the edge mode. 

\begin{figure}[htbp]
    \centering
    \begin{subfigure}[b]{0.48\textwidth}
\centering
\resizebox{0.6\linewidth}{!}{
\begin{tikzpicture}[line cap=round]

\def\R{2.2}
\def\bandwidth{12pt}
\def\thetaA{235}
\def\thetaB{325}
\def\thetaSym{305}

\draw[black!75, line width=0.9pt] (0,0) circle (\R);

\draw[densely dashed, gray!45]
    (0,-2.35) -- (0,2.35);

\draw[green!65!black, line width=\bandwidth,
      opacity=0.42]
    (108:\R) arc (108:72:\R);

\draw[blue!75, line width=\bandwidth,
      opacity=0.42]
    (250:\R) arc (250:220:\R);

\draw[orange!80, line width=\bandwidth,
      opacity=0.42]
    (340:\R) arc (340:310:\R);
\fill[blue!80!black] (\thetaA:\R) circle (1.5pt);
\fill[orange!80!black] (\thetaB:\R) circle (1.5pt);

\draw[gray!70, densely dashed]
    (0,0) -- (\thetaSym:2.45);
\draw[gray!75, fill=white]
    (\thetaSym:\R) circle (2pt);

\draw[orange!65!black, thin]
    (0,0) -- (\thetaB:1.85);

\draw[black, line width=0.9pt, <-]
    (\thetaB:0.95)
    arc (\thetaB:\thetaSym:0.95);
\node at (315:1.30) {$\alpha$};

\node[green!45!black] at (90:2.58) {$f$};
\node[blue!80!black] at (235:2.60) {$r_A$};
\node[orange!75!black] at (325:2.62) {$r_B$};

\end{tikzpicture}
}
\caption{}
\label{fig:asymmetric-bands}
\end{subfigure}
    \hfill
    \begin{subfigure}[b]{0.48\textwidth}
        \centering
        \includegraphics[width=\linewidth]{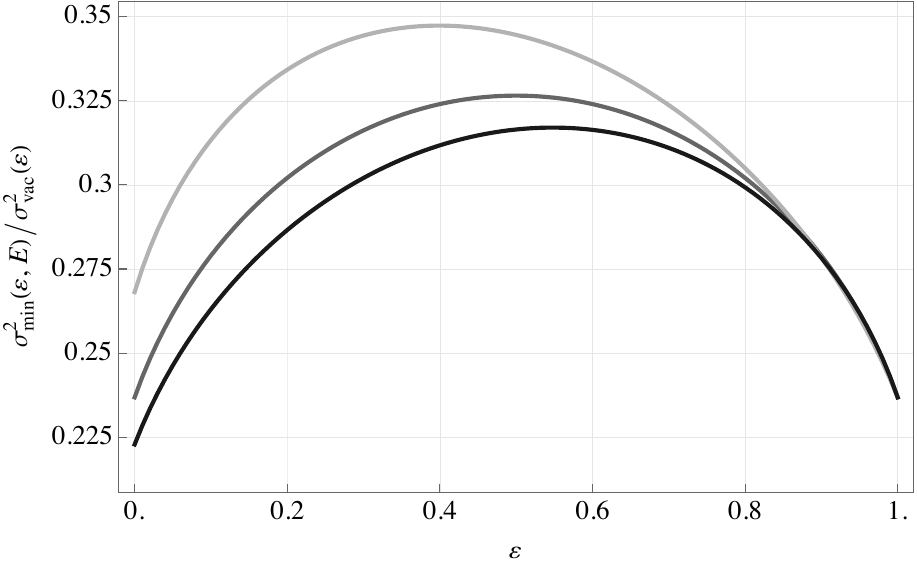}
        \caption{}
    \end{subfigure}
\caption{(a) Geometry of the two-party setup. A non-zero angle $\alpha$ introduces an asymmetry beyween the reference frames $A$ and $B$. (b) Fractional advantage of the optimal state in the two-party setup as a function of the importance dimensionless parameter $\epsilon$, for a choice of $\alpha/2\pi=\{0.1,0,-0.1\}$ from lighter to darker. There is an evident trade-off when optimising both records at the same time. The worst case-scenario, for the symmetric $\alpha=0$ configuration, corresponds to giving the same weight to both readings ($\epsilon=1/2$). Negative $\alpha$'s give an advantage as they push one reference region towards the antipode of the target one.}
    \label{fig:trade-off}
\end{figure}
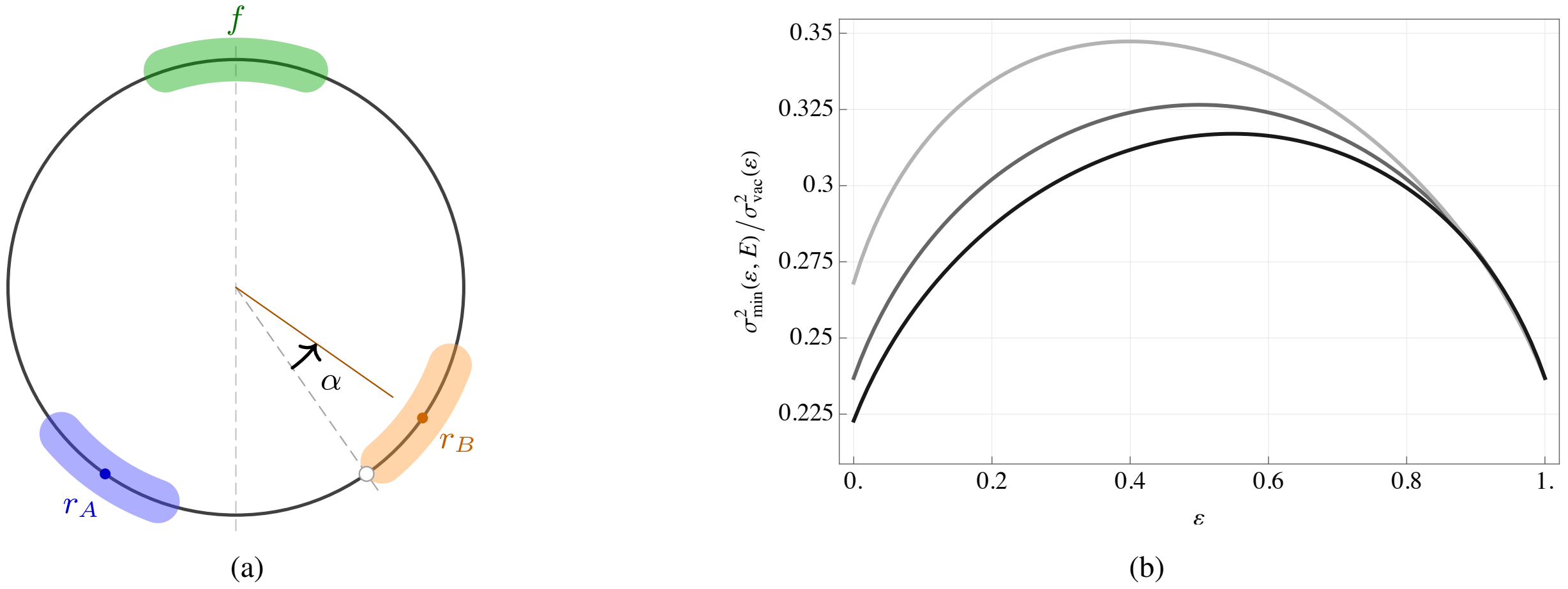

Let us see this in action. First, let us introduce a parameter $\epsilon\in[0,1]$ to weigh the importance of the sharpness of the observable relational to $A$ in the state preparation. In particular, the auxiliary Hamiltonian to minimise becomes
\begin{equation}
\hat H_{\lambda,\epsilon} =\hat H_{\mathrm{exc}} +\frac{\lambda}{2} \left[\epsilon \hat \Phi_{f|A,w}^2+(1-\epsilon)\hat \Phi_{f|B,w}^2\right] \, .
\label{eq:two-reference-auxiliary}
\end{equation}
The optimisation task is the same as before (optimal states are still multimode squeezed vacuum states), with the sole difference that
\begin{equation}
    K_\lambda \to \epsilon K_{\lambda}^{(A)} + (1-\epsilon) K_{\lambda}^{(B)} \, ,
\end{equation}
yielding
\begin{equation}
\sigma^2_{\min}(\epsilon,E) =\frac{\hbar}{2}\left[\epsilon\,d_A^{ T}K_{\lambda_\epsilon,\epsilon}^{-1/2}d_A +(1-\epsilon)\,d_B^{T}K_{\lambda_\epsilon,\epsilon}^{-1/2}d_B \right]\,.
\end{equation}
In the limits $\epsilon\to 0,1$ we trivially recover the single-reference result. However, if $A$ and $B$ are identical frames for the target observable (meaning that the result is invariant under exchange $\epsilon \to 1-\epsilon$) we can show that the balanced compromise always underperforms compared to using a single reference frame. Indeed, differentiating with respect to $\epsilon$ we obtain
\begin{equation}
    \frac{\partial \sigma^2_{\min}}{\partial\epsilon} =\sigma^2_{w,A}(\epsilon)-\sigma^2_{w,B}(\epsilon) \, ,
\end{equation}
meaning that (for identical frames) $\epsilon=1/2$ is required to achieve that maximum. The fact that it is a maximum, and not a minimum, follows trivially from concavity of the infimum. We see the trade-off in action in Figure~\ref{fig:trade-off}, where we numerically compute the minimal variance for the whole range of $\epsilon$, for two copies of the regional frame analysed in Figure~\ref{fig:squeezing}. We further numerically explore the effect of displacing the reference frame $B$ from the symmetric configuration, as a function of a displacement angle $\alpha$, as shown in the left panel of Figure~\ref{fig:trade-off}.

Optimising the balanced individual readings is similarly counterproductive if one is interested only in the combined relational observable $\sigma_{w,AB}$, since
\begin{equation}
   \frac{1}{2} (\sigma^2_{w,A}+\sigma^2_{w,B})= \sigma^2_{w,AB}+ \frac{1}{4}\mathrm{Var}(\hat{\Phi}_{A|B}) \geq \sigma^2_{w,AB} \, .
\end{equation}
Requiring both individual readings to be equally sharp, uses some of the fixed energy budget to localise the value of the edge-mode. If instead the individual records are neglected, and solely the sharpness of the relational field amplitude with respect to the combined $A\cup B$ region is optimised, there is a clear advantage. This comparison does not imply that a joint measurement improves over classical post-processing. Indeed, once the state has been optimised for the combined reading, its optimal variance can also be attained by jointly measuring the two commuting relational observables and averaging their outcomes, since the joint observable is simply a linear combination of the local ones.

\section{Conclusion}\label{sec:conclusion}
We constructed ideal regional quantum reference frames (QRFs) for the global shift symmetry of a real massless scalar field on a circle. The spatial discretisation of the field as a periodic chain of harmonic oscillators gives the construction a simple physical interpretation. The centre-of-mass displacement of a set of sites serves as the reference, and their regional momentum generates a uniform shift of those sites. 

The reduction to a regional QRF and the unitary transformations between regional charts are analogous in the discrete and continuum case. In the continuum case, the reference field amplitude can still be accessed entirely within the chosen region, but the conjugate momentum that generates a uniform shift within the region must extend beyond its boundary. This marks the main difference between the lattice model and the continuum field: the smearing profile of the conjugate momentum falls smoothly to zero across an external \emph{collar}~\cite{Fewster2016SplitProperty}, whose width can be arbitrarily small, but must remain positive. Closing the collar makes the vacuum fluctuations of the regional charge diverge logarithmically.

The need for a collar raises an algebraic question: which observables remain accessible when measurements are confined to the regional reference? We show that the algebra of observables strictly localised in a proper region cannot be a type~I factor. Such a factor would contain a bounded unitary operator that implements a sharp shift throughout that region, which the field representation does not admit.

For two complementary regions, the same question of accessibility reveals a connection between the relational edge mode and the extra particle~\cite{CastroRuizOreshkov2025RelativeSubsystems, GarmierHausmannCastroRuiz2025, SahdoCastroRuiz2026}. In the neutral sector, corresponding to the perspective-neutral approach to QRFs~\cite{VanrietveldeHohnGiacominiCastroRuiz2020, DeLaHametteEtAl2021} the two regional charges are equal and opposite. The charge of either region is conjugate, up to a sign, to the difference between their field averages. Together these observables form a relational edge mode. If the local accessible algebra is enlarged to include the regional charge, its value can be accessed, but invariant measurements confined to either region cannot detect coherences between different values of the charge. Our analysis suggests a more general connection between the extra-particle degrees of freedom~\cite{CastroRuizOreshkov2025RelativeSubsystems} and the relational edge mode: in our model, the value of the edge momentum provides the classical charge label of the Abelian extra-particle description. Retaining this label across sectors, fixing its value, or lacking access to it recovers the extra-particle, perspective-neutral, or operational approaches to QRFs, respectively.

It would be interesting to investigate how this connection extends to non-Abelian symmetries, both on the lattice and crucially in continuum field theory, where the extra-particle algebra can contain quantum degrees of freedom beyond a classical charge label. In gauge theories with local constraints, a related question is how these degrees of freedom are linked to the boundary charges that enter the regional observable algebras \cite{SahdoCastroRuiz2026,CarrozzaHohn2022,AraujoRegadoHohnSartini2025,SpalvieriGarmierGiacomini2025LOCCGauge}.

In our discussion, we have always assumed that different reference regions are disjoint. Allowing the averaging profiles to overlap raises an important question: which observables are accessible from both perspectives, and what does it mean to relate different overlapping perspectives? Comparing the common observable algebra with the spatial overlap of the profiles would give a concrete setting in which to study shared information across regional QRF perspectives~\cite{ValenteEtAl2026}.

Finally, we illustrate a physical situation in which our construction arises naturally. In particular, we study the task of determining how sharply one regional amplitude can be read relative to another. In a ring-shaped Bose-Einstein condensate, this relative amplitude corresponds to the difference between the average condensate phases in the target and reference regions.  For fixed target and reference profiles and a fixed mean excitation energy, the minimum variance of the relative amplitude is attained by a correlated multimode squeezed vacuum. A trade-off arises when the same preparation must provide precise readings relative to two different references. Bounds for profiles of prescribed width and separation could establish how much of this precision cost follows from regional localisation alone. These phase fluctuations also provide a link to analogue gravity because, at long wavelengths, they propagate as a massless scalar field on an effective spacetime geometry determined by the condensate background~\cite{Barcelo:2000tg}. It would be interesting to investigate this further. This metrology task also has a possible extension to temporal references and to linearised gravity~\cite{HohnSmithLock2021,Giacomini2021Spacetime,GiacominiBrukner2022}. In the latter setting, fluctuations in the separation of freely falling bodies depend on the state of the gravitational field~\cite{ParikhWilczekZahariade2021}. Quantum reference fields offer a way to formulate the corresponding relational question~\cite{ChenGiacomini2026ReferenceFields}. 

This work is the starting point to extend this treatment to more complex situations such as local gauge theories and gravity. The long-term goal of this research is to formulate diffeomorphisms relationally and without a fixed spacetime structure.

\subsection*{Note added}
While completing this manuscript, we became aware of independent work in which a construction similar to ours was developed \cite{AgulloMele2026}. That work examines the von Neumann entropy of subsystems and the mutual information between them, complementing our study of the precision of relative field readings at fixed energy. We look forward to exploring the connections between the two approaches further.

\acknowledgments{This work was carried out with the support of the Italian Ministry of University and Research (MUR) under the Funding for Research Projects (FIS 2), pursuant to Ministerial Decree No. 23314 of 11/12/2024 (Project Q-GraSp, FIS-2023-02629). This work was made possible through the support of the WOST, WithOut SpaceTime project (https://withoutspacetime.org), led by the Center for Spacetime and the Quantum (CSTQ), and supported by Grant ID\#~63683 from the John Templeton Foundation (JTF). The opinions expressed in this work are those of the author(s) and do not necessarily reflect the views of the John Templeton Foundation or those of the Center for Spacetime and the Quantum. This work has been supported by Klaus Tschira Foundation. F.G. also acknowledges support from the Swiss National Science Foundation via the Ambizione Grant PZ00P2-208885. This work contributes to the European Union COST Actions CA23130 \textit{Bridging high and low energies in search of quantum gravity} and CA23115 \textit{Relativistic Quantum Information}. GPT-Astra has been used to assist with some of the proofs and some of the figures.}

\bibliography{bibliography}

\addtocontents{toc}{\protect\endinput}
\appendix

\section{Subsystem algebras and Hilbert-space factorisation}
\label[appendix]{app:subsystem-algebras}

Following the operational algebraic approach to quantum subsystems~\cite{Zanardi2001,ZanardiLidarLloyd2004,Bianchi:2024NonAbelianSREE}, a subsystem $S$ is specified by the von Neumann algebra $\mathcal{A}_S \subseteq \mathcal{L}(\mathcal{H})$ generated by its accessible observables, together with its commutant $\mathcal{A}_{\bar{S}} \coloneqq \mathcal{A}_S^\prime$ and their common centre
\begin{equation}
\mathcal{Z}_S = \mathcal{A}_S \cap \mathcal{A}_{\bar{S}}\,.
\end{equation}
When $\mathcal{A}_S$ admits type-I factor representations within each central sector, simultaneously diagonalising the commuting central observables $\{Q_k\}$ in $\mathcal{Z}_S$ decomposes the Hilbert space into superselection sectors labelled by their joint eigenvalues $\mathbf{Q}$:
\begin{equation}
\mathcal{H} = \bigoplus_{\mathbf{Q}} \mathcal{H}_S^{(\mathbf{Q})} \otimes \mathcal{H}_{\bar{S}}^{(\mathbf{Q})}\;, \;\; \mathcal{A}_S = \bigoplus_{\mathbf{Q}} \mathcal{L}\big(\mathcal{H}_S^{(\mathbf{Q})}\big) \otimes \mathbb{I}_{\bar{S}}\,,
\label{eq:app-standard-decomposition}
\end{equation}
where $\bigoplus$ denotes a direct sum for discrete spectra and a direct integral for continuous spectra.

In the presence of a symmetry group $G$ acting unitarily via $U(g) = e^{i\alpha_a T^a}$, we define the symmetry algebra $\mathcal{A}_G \coloneqq \langle \{T^a\} \rangle$, the algebra of globally invariant (physical) observables $\mathcal{A}_\mathrm{phys} \coloneqq \{O \in \mathcal{L}(\mathcal{H}) \vcentcolon [O, T^a] = 0\; \forall a\}$, and the invariant subsystem algebra
\begin{equation}
\mathcal{A}_\mathrm{phys,S} \coloneqq \mathcal{A}_\mathrm{phys} \cap \mathcal{A}_S\,.
\end{equation}
Under the same sectorwise type-I assumptions, the kinematical Hilbert space admits a decomposition adapted simultaneously to the global symmetry and to the subsystem~\cite{Bianchi:2024NonAbelianSREE}:
\begin{equation}
\mathcal{H} = \bigoplus_{Q,R} \mathcal{H}_G^{(Q)} \otimes \mathcal{H}_\mathrm{phys,S}^{(R)} \otimes \mathcal{H}_\mathrm{phys,\bar{S}}^{(Q,R)}\,,
\label{eq:app-symmetry-subsystem-decomposition}
\end{equation}
where $Q$ labels the global symmetry sectors in $\mathcal{Z}_G = \mathcal{A}_G \cap \mathcal{A}_\mathrm{phys}$ and $R$ labels the central sectors of the invariant subsystem algebra in $\mathcal{Z}_\mathrm{phys,S} = \mathcal{A}_\mathrm{phys,S} \cap\mathcal{A}_\mathrm{phys,\bar{S}}$. The three tensor factors carry, respectively, the global symmetry (gauge) degrees of freedom, the invariant degrees of freedom accessible within $S$, and the remaining invariant degrees of freedom compatible with $Q$ and $R$.

For the discrete translation group on the $N$-site lattice discussed in \Cref{sec:lattice-construction}, $G = (\mathbb{R}, +)$ is Abelian with global generator $\hat{\pi}_T = \sum_{i=0}^{N-1} \hat{\pi}_i$, so $\mathcal{A}_G = \mathcal{Z}_G = \langle \hat{\pi}_T \rangle$ is labelled by the total momentum eigenvalue $Q \in \mathbb{R}$. For a subregion $A$ of $n_A$ sites, upon choosing a reference site $a_*$, the invariant regional algebra $\mathcal{A}_\mathrm{phys,A} = \langle \{\hat{\phi}_\mu - \hat{\phi}_{a_*},\, \hat{\pi}_\mu\}_{\mu\in A\setminus \{a_*\}} \rangle$ has centre $\mathcal{Z}_\mathrm{phys,A} = \langle \hat{\pi}_A \rangle$, labelled by the regional momentum $R = \pi_A \in \mathbb{R}$. Since the total momentum of the complementary region $\bar{A}$ is fixed by $\hat{\pi}_{\bar{A}} = \hat{\pi}_T - \hat{\pi}_A$, \Cref{eq:app-symmetry-subsystem-decomposition} specialises to
\begin{equation}
\mathcal{H}_{\mathrm{kin}} = \bigoplus_{Q,\pi_A} \mathcal{H}_G^{(Q)} \otimes \mathcal{H}_\mathrm{phys,A}^{(\pi_A)} \otimes \mathcal{H}_\mathrm{phys,\bar{A}}^{(Q-\pi_A)}\,,
\end{equation}
which,  upon restricting to a fixed global charge sector $Q$, reduces to 
\begin{equation}
\mathcal{H}_{\mathrm{phys}}^{(Q)} \simeq \bigoplus_{\pi_A} \mathcal{H}_A^{(\pi_A)} \otimes \mathcal{H}_{\bar{A}}^{(Q-\pi_A)}\;.
\label{eq:phys-edge-discrete_app}
\end{equation}

\section{Regularity of smeared operators}
\label[appendix]{app:regularity_of_smeared_operators}

We derive the vacuum-norm criteria used in the main text, establish the different ultraviolet behaviour of the sharp regional field average and charge, and show that smooth periodic profiles give finite vacuum fluctuations.
\paragraph{Oscillator vacuum norms.}
At fixed circumference $L$, the non-zero modes have wave numbers $k_n=2\pi n/L$ and frequencies $\omega_n=v|k_n|$. At fixed time, the oscillator fields have the expansions
\begin{equation}
\begin{aligned}
 \hat\Phi_{\mathrm{osc}}(y) &= \sum_{n\neq0}\sqrt{\frac{\hbar}{2L\omega_n}}\, \left( \hat a_n e^{ik_ny}+\hat a_n^\dagger e^{-ik_ny}\right)\,,\\
 \hat\Pi_{\mathrm{osc}}(y)
 &= -i\sum_{n\neq0}\sqrt{\frac{\hbar\omega_n}{2L}}\,\left(\hat a_n e^{ik_ny} -\hat a_n^\dagger e^{-ik_ny}\right)\,.
\end{aligned}
\end{equation}
For real periodic profiles $f$ and $g$, define
\begin{equation}
 f_n=\frac1L\int_0^L \dd y\,f(y)e^{-ik_ny}\,,\qquad g_n=\frac1L\int_0^L \dd y\,g(y)e^{-ik_ny}\,. \label{eq:fourier_coefficients}
\end{equation}
Smearing and acting on the oscillator vacuum gives
\begin{equation}
\begin{aligned}
 \hat\Phi_{\mathrm{osc}}(f)\ket{0_{\mathrm{osc}}}
 &= \sum_{n\neq0}\sqrt{\frac{\hbar L}{2\omega_n}}\,, f_n\,\hat a_n^\dagger\ket{0_{\mathrm{osc}}}\,,\\
 \hat\Pi_{\mathrm{osc}}(g)\ket{0_{\mathrm{osc}}}
 &= i\sum_{n\neq0}\sqrt{\frac{\hbar L\omega_n}{2}}\,, g_n\,\hat a_n^\dagger\ket{0_{\mathrm{osc}}}\,.
\end{aligned}
\end{equation}
These expressions define vectors in Fock space whenever their coefficients are square summable. Orthogonality of the one-particle states then yields
\begin{equation}
\begin{aligned}
 \bigl\|\hat\Phi_{\mathrm{osc}}(f)\ket{0_{\mathrm{osc}}}\bigr\|^2
 &= \frac{\hbar L}{2v}\sum_{n\neq0}\frac{|f_n|^2}{|k_n|}\,,\\
 \bigl\|\hat\Pi_{\mathrm{osc}}(g)\ket{0_{\mathrm{osc}}}\bigr\|^2
 &= \frac{\hbar Lv}{2}\sum_{n\neq0}|k_n|\,|g_n|^2\,.
    \label{eq:field_norm}
\end{aligned}
\end{equation}

\paragraph{Sharp regional field average.}
Consider an interval $A=[a,b]$ of length $0<l_A=b-a<L$. The non-zero Fourier coefficients of its characteristic function are
\begin{equation}
 (\chi_A)_n= \frac1L\int_a^b \dd y\,e^{-ik_ny}
 = e^{-ik_n(a+b)/2}\frac{\sin(\pi n l_A/L)}{\pi n}\,,\qquad n\neq0\,.
\end{equation}
The regional field average uses the normalised profile $f=\chi_A/l_A$. Combining the contributions from positive and negative Fourier indices gives
\begin{equation}
 \left\|\hat\Phi_{\mathrm{osc}}\!\left(\frac{\chi_A}{l_A}\right)\ket{0_{\mathrm{osc}}}\right\|^2=\frac{\hbar L^2}{2\pi^3v l_A^2}\sum_{n=1}^{\infty}\frac{\sin^2(\pi n l_A/L)}{n^3}\,. \label{eq:partial_sum}
\end{equation}
Since $\sin^2(\pi n l_A/L)\leq1$, the series is bounded by $\sum_{n\geq1}n^{-3}$ and therefore converges. The sharp regional field average has finite oscillator vacuum variance.

\paragraph{Sharp regional charge.}
For the sharp momentum profile $g=\chi_A$, the corresponding vacuum-norm series is
\begin{equation}
 \bigl\|\hat\Pi_{\mathrm{osc}}(g)\ket{0_{\mathrm{osc}}}\bigr\|^2=\frac{2\hbar v}{\pi}\sum_{n=1}^{\infty}\frac{\sin^2(\pi n l_A/L)}{n}\,. \label{eq:log_divergence_der}
\end{equation}
To determine its convergence, we examine its partial sums. Writing $\theta=\pi l_A/L\in(0,\pi)$ and using $\sin^2x=(1-\cos2x)/2$, we obtain
\begin{equation}
\begin{aligned}
 S_M
 &= \frac{2\hbar v}{\pi}
    \sum_{n=1}^{M}\frac{\sin^2(n\theta)}{n}\\
 &= \frac{\hbar v}{\pi}\left[\sum_{n=1}^{M}\frac1n -\sum_{n=1}^{M}\frac{\cos(2n\theta)}{n}
    \right]\,.
\end{aligned}
\end{equation}
For the harmonic sum, we use the standard asymptotic expansion
\begin{equation}
 \sum_{n=1}^{M}\frac1n= \log M+\gamma_{\mathrm E}+O(M^{-1})\,,
\end{equation}
where $\gamma_{\mathrm E}$ is the Euler-Mascheroni constant.

To study the oscillatory contribution, write $\cos(2n\theta)=\operatorname{Re}(e^{2in\theta})$. The geometric-series identity gives
\begin{equation}
 \sum_{n=1}^{M}e^{2in\theta}= e^{2i\theta}\frac{1-e^{2iM\theta}}{1-e^{2i\theta}}\,.
\end{equation}
Using $1-e^{2i\theta}=-2i\,e^{i\theta}\sin\theta$, we find
\begin{equation}
 \left|\sum_{n=1}^{M}e^{2in\theta}\right|= \frac{|1-e^{2iM\theta}|}{2|\sin\theta|}\leq \frac{1}{|\sin\theta|}\,,
\end{equation}
where the inequality follows from $|1-e^{2iM\theta}|\leq1+|e^{2iM\theta}|=2$. For each fixed $\theta\in(0,\pi)$, this bound is finite and independent of $M$. Consequently,
\begin{equation}
 \left|\sum_{n=1}^{M}\cos(2n\theta)\right|\leq \left|\sum_{n=1}^{M}e^{2in\theta}\right|\leq \frac{1}{|\sin\theta|}\,.
\end{equation}
The cosine partial sums are therefore bounded. Since $1/n$ decreases to zero, the Dirichlet test gives convergence of\\$\sum_{n\geq1}\cos(2n\theta)/n$. Combining this result with the harmonic-sum asymptotics yields
\begin{equation}
 S_M=\frac{\hbar v}{\pi}\log M+O(1)\qquad (M\to\infty)\,. \label{eq:log_divergence}
\end{equation}
Here the remainder $O(1)$ stays bounded at fixed $L$ and $l_A$. It contains the finite correction to the harmonic sum and the convergent oscillatory contribution. The partial sums $S_M$ therefore grow without bound, establishing the logarithmic divergence of the sharp regional charge's vacuum fluctuations. This divergence involves only non-zero modes and persists after reduction at $C=0$.

\paragraph{Smooth periodic profiles.}
For a $C^2$ profile $g$ on the circle, integrating by parts twice gives
\begin{equation}
 g_n= -\frac{1}{Lk_n^2}\int_0^L \dd y\,g^{\prime \prime}(y)e^{-ik_ny}\,,\qquad n\neq0\,.
\end{equation}
The boundary terms vanish because $g$ and $g^\prime$ are periodic and $e^{-ik_nL}=1$. Consequently,
\begin{equation}
\begin{aligned}
 |g_n|
 &\leq \frac{\|g^{\prime\prime}\|_{L^1}}{L|k_n|^2}\,,\\
 \sum_{n\neq0}|k_n|\,|g_n|^2
 &\leq \frac{\|g^{\prime\prime}\|_{L^1}^2}{L^2}\sum_{n\neq0}\frac1{|k_n|^3}<\infty\,.
\end{aligned}
\end{equation}
The same estimate applied to a $C^2$ field profile $f$ gives
\begin{equation}
 \sum_{n\neq0}\frac{|f_n|^2}{|k_n|}\leq \frac{\|f^{\prime\prime}\|_{L^1}^2}{L^2}
      \sum_{n\neq0}\frac1{|k_n|^5}
 <\infty\,.
\end{equation}
Thus $C^2$ periodic profiles already suffice for finite vacuum norms of both smeared observables. For the $C^\infty$ periodic profiles used in our construction, repeated integration by parts gives Fourier decay faster than any inverse power of $|k_n|$. At fixed $L$, the non-zero wave numbers satisfy $|k_n|\geq2\pi/L>0$.

\subsection{Explicit smearing profiles}
\label[appendix]{app:explicit_regional_profiles}
We give the explicit profiles used in \cref{fig:three-regional-profiles} and verify their properties. The construction uses the normalised bump and the smooth step introduced in the main text.

\paragraph{Reference profiles and regional supports.}
Choose three equally spaced reference centres,
\begin{equation}
    x_A=\frac{L}{6}\,,\quad x_B=\frac{L}{2}\,, \quad x_C=\frac{5L}{6}\,,
\end{equation}
and a common regional width $0<w<L/3$. Before periodic extension, the normalised reference bump is
\begin{equation}
    b_{\alpha,w}(x)=
    \begin{cases}
        Z_w^{-1}\exp\!\bigl[-\tfrac{1}{1-u_{\alpha,w}(x)^2}\bigr]\,,
        & |u_{\alpha,w}(x)|<1\,,\\[4pt]
        0\,, & |u_{\alpha,w}(x)|\geq1\,,
    \end{cases}
\end{equation}
where $u_{\alpha,w}(x)\coloneqq2(x-x_\alpha)/w$ and
\begin{equation}
    Z_w=\int_{x_\alpha-w/2}^{x_\alpha+w/2}
    \dd x\,\exp\!\bigl[-\tfrac{1}{1-4(x-x_\alpha)^2/w^2}\bigr]
    =\frac{w}{2}\int_{-1}^{1} \dd u\,e^{-1/(1-u^2)}\,.
\end{equation}
Thus $Z_w$ includes the spatial rescaling factor. We define the periodic reference profile by
\begin{equation}
    r_{\alpha,w}(x)=\sum_{n\in\mathbb Z}b_{\alpha,w}(x-nL)\,.
\end{equation}
Each profile is non-negative, smooth, and normalised on the circle,
\begin{equation}
    \int_0^L \dd x\,r_{\alpha,w}(x)=1\,.
\end{equation}
Its support defines the reference region,
\begin{equation}
    A_\alpha=\operatorname{supp}r_{\alpha,w} =\left[x_\alpha-\frac{w}{2},\,x_\alpha+\frac{w}{2}\right]\,,
\end{equation}
with the interval understood periodically. The regions are mutually disjoint.

\paragraph{Charge profiles and collars.}
The gap between neighboring reference regions has width
\begin{equation}
    \ell=\frac{L}{3}-w>0\,.
\end{equation}
We choose this gap as the width of each charge transition. The periodic charge profiles are
\begin{equation}
    g_{\alpha,w,\ell}(x)= \sum_{n\in\mathbb Z} S\!\left(\frac{x-x_\alpha-nL+w/2+\ell}{\ell}\right) S\!\left(\frac{x_\alpha+nL+w/2+\ell-x}{\ell}\right)\,.
    \label{eq:explicit_three_charge_profiles}
\end{equation}
Each profile equals one throughout $A_\alpha$ and begins to decrease immediately outside its boundary. It reaches zero at the boundary of the neighbouring region. Its support is
\begin{equation}
    A_\alpha^+=\operatorname{supp}g_{\alpha,w,\ell} =\left[x_\alpha-\frac{w}{2}-\ell,\,x_\alpha+\frac{w}{2}+\ell\right]\,.
\end{equation}
The collar $\mathcal C_\alpha=A_\alpha^+\setminus A_\alpha$ therefore has two portions, each of width $\ell$. Between neighboring regions, the corresponding collar portions occupy the same gap.

For the example in \cref{fig:three-regional-profiles}, we choose
\begin{equation}
    w=\frac{L}{6}\,,\quad \ell=\frac{L}{6}\,.
\end{equation}
The reference supports have half-width $L/12$, while the charge supports have half-width $L/4$.

\paragraph{Partition of unity.}
Consider neighbouring centres $x_\alpha$ and $x_\beta=x_\alpha+L/3$, using unwrapped coordinates when crossing the periodic boundary. Write
\begin{equation}
    a=x_\alpha+\frac{w}{2}\,,\quad b=x_\beta-\frac{w}{2}=a+\ell
\end{equation}
for the two boundaries enclosing their gap. For $a\leq x\leq b$, define $t\coloneqq(x-a)/\ell\in[0,1]$. The right-hand transition of $g_{\alpha,w,\ell}$ and the left-hand transition of $g_{\beta,w,\ell}$ then have arguments
\begin{equation}
    \frac{x_\alpha+w/2+\ell-x}{\ell}=1-t\,, \quad \frac{x-x_\beta+w/2+\ell}{\ell}=t\,.
\end{equation}
The remaining factor in each charge profile equals one throughout the gap, while the third charge profile vanishes there. Hence
\begin{equation}
    g_{\alpha,w,\ell}(x)=S(1-t)\,, \quad g_{\beta,w,\ell}(x)=S(t).
\end{equation}
From the explicit definition of the smooth step,
\begin{equation}
    S(1-t)=1-S(t)\,,
\end{equation}
including the endpoints. Consequently,
\begin{equation}
    g_{\alpha,w,\ell}(x)+g_{\beta,w,\ell}(x) =[1-S(t)]+S(t)=1\,.
\end{equation}
Within each reference region, its own charge profile equals one and the others vanish. Applying the same argument to all three gaps, including the one across $0\equiv L$, gives
\begin{equation}
    g_{A,w,\ell}(x)+g_{B,w,\ell}(x)+g_{C,w,\ell}(x)=1 \qquad\text{for all }x\,.
    \label{eq:examples_charge_partition}
\end{equation}

\paragraph{Duality and regularity.}
Each charge profile equals one on the support of its own reference profile and vanishes on the supports of the other references:
\begin{equation}
    g_{\beta,w,\ell}(x)=\delta_{\alpha\beta} \qquad\text{for }x\in A_\alpha\,.
\end{equation}
Normalisation therefore gives
\begin{equation}
    \int_0^L \dd x\,r_{\alpha,w}(x)g_{\beta,w,\ell}(x) =\delta_{\alpha\beta}\int_0^L \dd x\,r_{\alpha,w}(x) =\delta_{\alpha\beta}\,.
\end{equation}
All profiles are non-negative, smooth, and periodic. Their Fourier coefficients consequently decay rapidly, ensuring finite oscillator vacuum norms for both field and momentum smearings. The construction therefore satisfies positivity, regularity, duality, and the partition of unity, with the reference regions and their collars determined by the corresponding supports.

\section{Explicit construction of the canonical complement}\label{app:canonical_complement_construction}
We saw that a field reading relative to $A$ can still contain an offset between two regional references. We now separate these offsets from the rest of a general relative reading $\mathcal R_A[\hat\Phi(f)]$. To do so using the relativisation map, we resolve the reading into contributions that can each be referred to a particular regional mean. The charge profiles provide such a decomposition because they form a partition of unity,
\begin{equation}
    f_\alpha(x)\coloneqq f(x)g_\alpha(x)\,,\qquad f=\sum_\alpha f_\alpha\,, \qquad  a_\alpha(f)\coloneqq\int_0^L f_\alpha(x)\,\mathrm{d}x\, .
    \label{eq:regional-smearing-decomposition}
\end{equation}
Consider the contribution $\hat\Phi(f_\alpha)$. We compare the reading obtained using $\hat\Phi_A$ as the origin with the one obtained using $\hat\Phi_\alpha$. In both cases, the relativisation map \eqref{eq:relativization_map} multiplies the reference amplitude by the integral of the smearing, which for $f_\alpha$ is precisely $a_\alpha(f)$. Hence
\begin{equation}
\begin{aligned}
\mathcal R_A[\hat\Phi(f_\alpha)]
&=\hat\Phi(f_\alpha)-a_\alpha(f)\hat\Phi_A\,,\\
\mathcal R_\alpha[\hat\Phi(f_\alpha)]
&=\hat\Phi(f_\alpha)-a_\alpha(f)\hat\Phi_\alpha\,.
\end{aligned}
\label{eq:two-regional-readings}
\end{equation}
Changing the origin from $\alpha$ to $A$ therefore adds $a_\alpha(f)(\hat\Phi_\alpha-\hat\Phi_A)$ to this contribution:
\begin{equation}
        \mathcal R_A[\hat\Phi(f_\alpha)] = \underbrace{\bigl[\hat\Phi(f_\alpha) -a_\alpha(f)\hat\Phi_\alpha\bigr]}_{\mathcal R_\alpha[\hat\Phi(f_\alpha)]} \quad+ \quad a_\alpha(f)\bigl(\hat\Phi_\alpha - \hat\Phi_A\bigr)\,.
\end{equation}
then summing over $\alpha$ gives
\begin{equation}
    \mathcal R_A[\hat\Phi(f)] =\sum_\alpha \mathcal R_A[\hat\Phi(f_\alpha)] = \underbrace{\sum_\alpha\mathcal R_\alpha[\hat\Phi(f_\alpha)]}_{\displaystyle \hat\Phi_\perp(f)}\qquad+\underbrace{\sum_\alpha a_\alpha(f)(\hat\Phi_\alpha-\hat\Phi_A)}_{\displaystyle \text{reference offsets relative to }A}.
\label{eq:relative-reading-decomposition}
\end{equation}
The sum of reference offsets contains all the dependence on the choice of $A$. Substituting \eqref{eq:two-regional-readings} into the sum labelled $\hat\Phi_\perp(f)$ gives
\begin{equation}
\hat\Phi_\perp(f)=\hat\Phi(f)-\sum_\alpha a_\alpha(f)\hat\Phi_\alpha \,, \qquad a_\alpha(f)=\int_0^L \dd x\, f(x) g_\alpha(x)\,.
\label{eq:complementary-field-reading_app}
\end{equation}
The coefficient $a_\alpha(f)$ is fixed by
\begin{equation}
[\hat\Phi_\perp(f),\hat \Pi_\alpha]=i\hbar\int_0^L \dd y\,f(y)g_\alpha(y)-a_\alpha(f)=0\,.
\end{equation}
To see why the subtraction involves every regional mean, consider a shift of amplitude $\epsilon$ generated by one regional charge $\hat\Pi_\alpha$. The smeared field transforms as
\begin{equation}
    \hat\Phi(f)\longmapsto\hat\Phi(f)+\epsilon a_\alpha(f)\,\mathds{1}\,.
\label{eq:regional-shift-smeared-field}
\end{equation}
By regional duality, the same shift acts on the reference amplitudes as
\begin{equation}
    \hat\Phi_\alpha \longmapsto \hat\Phi_\alpha +\epsilon \left(\int_0^L \dd x\, r_\alpha(x)g_\beta(x)\right)\mathds{1}=\hat\Phi_\alpha+\epsilon\delta_{\alpha\beta}\,\mathds{1}\,.
    \label{eq:regional-shift-reference-amplitude}
\end{equation}
Thus, the term $a_\alpha(f)\hat\Phi_\alpha$ subtracted in \eqref{eq:complementary-field-reading_app} changes by exactly the same amount as $\hat\Phi(f)$. Since equal-time field amplitudes commute, $\hat\Phi_\perp(f)$ also commutes with every $\hat\Phi_\alpha$. For a normalised profile $f$ supported in $A_\alpha$, \eqref{eq:complementary-field-reading_app} reduces to $\hat\Phi_\perp(f)=\hat\Phi(f)-\hat\Phi_\alpha$. This measures the difference between the field amplitude sampled by $f$ and the mean amplitude of the same region. In this sense it plays the same role as a lattice bulk coordinate: it describes an amplitude difference within the region that is unchanged by a uniform shift of the region
as a whole.

We apply the same criterion to the momentum observables. Subtracting the collective momenta that account for the action of $\hat\Pi(h)$ on the regional amplitudes gives
\begin{equation}
    \hat\Pi_\perp(h)=\hat\Pi(h)-\sum_\alpha b_\alpha(h)\hat\Pi_\alpha\,,\qquad b_\alpha(h)=\int_0^L \dd x \, r_\alpha(x)h(x)\,,
    \label{eq:complementary-momentum-reading_app}
\end{equation}
and the coefficient $b_\alpha(h)$ is determined by
\begin{equation}
[\hat \Phi_\alpha,\hat\Pi_\perp(h)]=i\hbar\int_0^L \dd y\,r_\alpha(y)h(y)-b_\alpha(h) ,=0.
\end{equation}
The observables $\hat\Phi_\perp(f)$ and $\hat\Pi_\perp(h)$ describe the degrees of freedom we sought. They form a \emph{complement} because adding back the collective contributions in \eqref{eq:complementary-field-reading_app} and \eqref{eq:complementary-momentum-reading_app} reconstructs every smeared field and momentum observable. The separation is \emph{canonical} because each of these observables commutes with both members of every regional canonical pair. We therefore call the modes they describe the \emph{canonical complement} of the regional reference modes. 

\section{Regional algebras are not of type I}\label{app:regional_algebras_are_not_of_type_I}
Here we prove that the regional algebra $\mathcal A_A$, defined in \cref{eq:regional-algebra-qft-definition}, is not a type-I factor. Suppose, by contradiction, that $\mathcal A_A$ were a type-I factor, so that $\mathcal A_A=\mathcal B(\mathcal H_A)\otimes\mathds{1}_{\bar A}$. The global shift $\hat G(c)=e^{ic\hat\Pi_0/\hbar}$ preserves the regional algebra and thus defines an automorphism of $\mathcal A_A$. Since every automorphism of a type-I factor is inner, there must be a unitary $\hat V_A(c)\in\mathcal A_A$ implementing the shift on observables supported in $A$. By locality, it leaves observables supported in $\bar A$ unchanged. It would therefore implement the sharp regional shift
\begin{equation}
    \hat\Phi(x)\longmapsto \hat\Phi(x)+c\,\chi_A(x)\,\mathds{1}\,,
\end{equation}
where $\chi_A$ is the characteristic function of $A$. On smeared fields the shift acts as
\begin{equation}
\begin{aligned}
    \hat\Phi(f) &\longmapsto \hat\Phi(f)+\lambda_A(f)\,\mathds{1}\,,\\
    \lambda_A(f)
    &=c\int_0^L \dd x\,\chi_A(x)f(x) =c\int_A \dd x f(x)\,.
\end{aligned}
\end{equation}
Such a coherent displacement is unitarily implementable in the oscillator Fock representation if and only if $\lambda_A$ is bounded with respect to the one-particle norm \cite{DerezinskiGerard}. In particular, there must exist a finite constant $C$, independent of $f$, such that
\begin{equation}
    \abs{\lambda_A(f)}^2\leq C\|f\|_\Phi^2\,.
\end{equation}
Thus, to rule out $\hat V_A(c)$ for $c\neq0$, it suffices to show that this bound fails. It suffices to test this bound on zero-mean smearings, for which the zero-mode contribution vanishes. With the Fourier convention of \cref{eq:fourier_coefficients}, the required norm is already given in \cref{eq:field_norm}:
\begin{equation}
    \|f\|_\Phi^2 \coloneqq\bigl\|\hat\Phi_{\mathrm{osc}}(f)\ket{0_{\mathrm{osc}}}\bigr\|^2 =\frac{\hbar L}{2v}\sum_{n\neq0}\frac{|f_n|^2}{|k_n|}\,.
\end{equation}

To test the bound for $c\neq0$, choose the real, smooth, zero-mean smearing $f_M$ with Fourier coefficients
\begin{equation}
    (f_M)_n=
    \begin{cases}
        |k_n|(\chi_A)_n\,, & 0<|n|\leq M\,,\\ 0\,,              & \text{otherwise}\,.
    \end{cases}
\end{equation}
Define
\begin{equation}
    T_M=\sum_{0<|n|\leq M}|k_n|\,|(\chi_A)_n|^2\,.
\end{equation}
Then
\begin{equation}
\begin{aligned}
    \lambda_A(f_M)
    &=cL\sum_{0<|n|\leq M}(f_M)_n\,\overline{(\chi_A)_n}=cL T_M\,,\\
    \|f_M\|_\Phi^2
    &=\frac{\hbar L}{2v}T_M\,.
\end{aligned}
\end{equation}
Substituting into the required bound gives
\begin{equation}
    C\geq
    \frac{|\lambda_A(f_M)|^2}{\|f_M\|_\Phi^2}=\frac{2c^2Lv}{\hbar}T_M=\frac{4c^2}{\hbar^2}S_M\,,
\end{equation}
where $S_M=(\hbar Lv/2)T_M$ is precisely the sharp-charge partial sum defined in \cref{eq:partial_sum}. \Cref{eq:log_divergence} already establishes that $S_M=(\hbar v/\pi)\log M+O(1)$ at fixed $L$ and $0<l_A<L$. Thus no finite constant $C$ can satisfy the bound for all $M$, and no unitary $\hat V_A(c)\in\mathcal A_A$ can implement the required sharp regional shift. This contradicts the type-I factor assumption.

The ultraviolet obstruction comes from the discontinuity of the shift at the boundary. Smoothing the profile through a collar removes this divergence, while extending the action of the shift outside $A$.

\section{Finding the multimode squeezed vacum}\label{app:optimal_state}
We derive the state that minimises the variance of the relative field amplitude $\hat\Phi_{f|A,w}$ at fixed mean excitation energy. Throughout this appendix we retain the first $N$ real, non-zero Fourier modes. Only at the end we show what happens once the cutoff is removed. With this convention, the mode expansion of \cref{eq:expansion-application} reads
\begin{equation}
    \hat\Phi_{f|A,w}
    = \sum_{j=1}^{N} d_j(w)\hat q_j\,,
    \quad
    d_j(w) = \int_0^L \dd x\, h_w(x)e_j(x)\,,
    \label{eq:opt_relative}
\end{equation}
where $h_w=f-r_{A,w}$. We denote the column of these coefficients by $d_N$ and the diagonal matrix of original mode frequencies by $\Omega_N=\operatorname{diag}(\omega_1,\ldots,\omega_N)$.

\subsection{Energy constraint and auxiliary Hamiltonian} 
A phase-space displacement can set the first moments of any state $\rho$ to zero while preserving the variance of $\hat\Phi_{f|A,w}$. If $\rho_{\mathrm c}$ denotes the centred state, the energy saved by this displacement is
\begin{equation}
    \expval{\hat H_{\mathrm{exc}}}_\rho
    - \expval{\hat H_{\mathrm{exc}}}_{\rho_{\mathrm c}}= \frac12\sum_{j=1}^{N}\left(\expval{\hat p_j}_\rho^2+ \omega_j^2\expval{\hat q_j}_\rho^2\right)\,.
    \label{eq:opt_centring}
\end{equation}
We may therefore restrict the optimisation to centred states, for which the variance equals $\expval{\hat\Phi_{f|A,w}^{\,2}}_\rho$. Following \cref{eq:lagrange_multiplaier}, we impose the energy constraint with a Lagrange multiplier $\mu$. For $\mu>0$, setting $\lambda=1/\mu$ gives
\begin{equation}
    \begin{aligned}
        \mathcal L_\mu(\rho)
        &=\frac12\langle\hat\Phi_{f|A,w}^{\,2}\rangle_\rho + \mu\bigl(\expval{\hat H_{\mathrm{exc}}}_\rho-E\bigr)\\
        &= \mu\bigl(\expval{\hat H_{\lambda,N}}_\rho-E\bigr)\,,
    \end{aligned}
\end{equation}
where
\begin{equation}
    \begin{aligned}
        \hat H_{\lambda,N}
        &= \hat H_{\mathrm{exc}}+ \frac{\lambda}{2}\hat\Phi_{f|A,w}^{\,2}\\
        &= \frac12\hat p^T\hat p+ \frac12\hat q^T K_{\lambda,N}\hat q - \frac{\hbar}{2}\operatorname{Tr}\Omega_N\,, \quad K_{\lambda,N}=\Omega_N^2+\lambda d_Nd_N^T\,.
    \end{aligned}
    \label{eq:opt_auxiliary}
\end{equation}
At fixed $\mu$, minimising $\mathcal L_\mu$ thus amounts to finding the ground state of $\hat H_{\lambda,N}$. The value of $\lambda$ is then selected by the physical energy budget.

\subsection{Ground state of the auxiliary Hamiltonian}
For every non-zero $u\in\mathbb R^N$,
\begin{equation}
    u^T K_{\lambda,N}u = \sum_{j=1}^{N}\omega_j^2u_j^2 + \lambda(d_N^Tu)^2 > 0\,.
\end{equation}
Since $K_{\lambda,N}$ is real and symmetric and we just showed that is positive defined, its eigenvalues of $K_{\lambda,N}$ can therefore be written as $\nu_\ell^2$ with $\nu_\ell>0$. Choose an orthogonal matrix $O$ such that
\begin{equation}
    O^T K_{\lambda,N}O = \operatorname{diag}(\nu_1^2,\ldots,\nu_N^2)\,.
\end{equation}
Rotating the coordinates and momenta together, $\hat Q=O^T\hat q$ and $\hat P=O^T\hat p$, preserves the canonical commutators, since
\begin{equation}
    [\hat Q_\ell,\hat P_m] = i\hbar(O^TO)_{\ell m} = i\hbar\delta_{\ell m}\,.
\end{equation}
In these coordinates the auxiliary Hamiltonian is a sum of independent oscillators,
\begin{equation}
    \hat H_{\lambda,N} = \sum_{\ell=1}^{N} \left(\frac{\hat P_\ell^2}{2}+ \frac{\nu_\ell^2\hat Q_\ell^2}{2}\right)- \frac{\hbar}{2}\sum_{j=1}^{N}\omega_j\,.
\end{equation}
Each excitation of the $\ell$-th oscillator raises the auxiliary energy by $\hbar\nu_\ell$. The unique ground state is therefore the product of the vacua of these normal modes. We denote this state by $\Psi_{\lambda,N}$ and the corresponding lowest eigenvalue of $\hat H_{\lambda,N}$ by $e_N(\lambda)$. To see how this state solves the constrained problem, consider any centred state $\rho$ with $\expval{\hat H_{\mathrm{exc}}}_\rho\leq E$. ~\cref{eq:opt_auxiliary} gives
\begin{equation}
    \begin{aligned}
        \frac{\lambda}{2}\operatorname{Var}_\rho(\hat\Phi_{f|A,w})
        &=
        \expval{\hat H_{\lambda,N}}_\rho
        - \expval{\hat H_{\mathrm{exc}}}_\rho\\
        &\geq e_N(\lambda)-E\,.
    \end{aligned}
    \label{eq:opt_variational}
\end{equation}
The auxiliary ground state attains equality when its physical excitation energy equals $E$. 

\subsection{Gaussian wave-function and squeezing}
In the normal coordinates $Q_\ell$, each oscillator ground-state wave-function is proportional to $\exp[-\nu_\ell Q_\ell^2/(2\hbar)]$. Taking their product and returning to the original Fourier coordinates gives
\begin{equation}
    \begin{aligned}
        \Psi_{\lambda,N}(q)
        &= \frac{(\det B_{\lambda,N})^{1/4}}{(\pi\hbar)^{N/4}} \exp\!\left[-\frac{1}{2\hbar}q^T B_{\lambda,N}q\right]\,,\\
        B_{\lambda,N}
        &= O\,\operatorname{diag}(\nu_1,\ldots,\nu_N)O^T = K_{\lambda,N}^{1/2}\,.
    \end{aligned}
    \label{eq:opt_wavefunction}
\end{equation}
The square root appears because the oscillator frequencies $\nu_\ell$ determine the widths of their ground-state wave-functions, while the quadratic potential contains their squares. The optimisation over all field states has thus selected a centred pure Gaussian state. In this product state the normal modes are independent, with oscillator covariances
\begin{equation}
    \expval{\hat Q_\ell\hat Q_m}_\lambda = \frac{\hbar}{2\nu_\ell}\delta_{\ell m}\,, \quad \expval{\hat P_\ell\hat P_m}_\lambda = \frac{\hbar\nu_\ell}{2}\delta_{\ell m}\,.
\end{equation}
Their symmetrised position-momentum covariances vanish. Returning to the original Fourier modes through $\hat q=O\hat Q$ and $\hat p=O\hat P$ gives
\begin{equation}
    \begin{aligned}
        \Gamma_{qq}^{(\lambda,N)}
        &= \frac{\hbar}{2}O\,\operatorname{diag}(\nu_\ell^{-1})O^T = \frac{\hbar}{2}B_{\lambda,N}^{-1}\,,\\
        \Gamma_{pp}^{(\lambda,N)}
        &= \frac{\hbar}{2}O\,\operatorname{diag}(\nu_\ell)O^T
         = \frac{\hbar}{2}B_{\lambda,N}\,,\\
        \Gamma_{qp}^{(\lambda,N)} &= 0\,.
    \end{aligned}
    \label{eq:opt_covariances}
\end{equation}
To identify the quadratures in which the squeezing is diagonal, we compare the auxiliary ground-state wave-function in \cref{eq:opt_wavefunction} with the original field vacuum. We introduce the dimensionless canonical quadratures
\begin{equation}
    \hat x=\frac{\Omega_N^{1/2}\hat q}{\sqrt{\hbar}},
    \quad
    \hat y=\frac{\Omega_N^{-1/2}\hat p}{\sqrt{\hbar}}\,.
\end{equation}
Substituting $q=\sqrt{\hbar}\,\Omega_N^{-1/2}x$ into
\cref{eq:opt_wavefunction} gives
\begin{equation}
    \psi_{\lambda,N}(x)
    \propto
    \exp\!\left[
        -\frac12 x^T
        \bigl(\Omega_N^{-1/2}B_{\lambda,N}\Omega_N^{-1/2}\bigr)x
    \right]\,.
\end{equation}
For the original vacuum, $B_{0,N}=\Omega_N$, so the matrix in parentheses reduces to the identity. The vacuum wave-function is therefore proportional to $\exp(-x^Tx/2)$, with variance $1/2$ in each dimensionless quadrature.

To diagonalise the auxiliary Gaussian, we use the normal-mode frequencies $\nu_m$ and the orthogonal matrix $O$ obtained above. Since $B_{\lambda,N}=K_{\lambda,N}^{1/2}=O\,\operatorname{diag}(\nu_1,\ldots,\nu_N)O^T$, the entries of the matrix in the exponent are
\begin{equation}
    \bigl(\Omega_N^{-1/2}B_{\lambda,N}\Omega_N^{-1/2}\bigr)_{jk}
    =
    \frac{1}{\sqrt{\omega_j\omega_k}}
    \sum_{m=1}^{N} O_{jm}\,\nu_m\,O_{km}\,.
\end{equation}
We denote the eigenvalues of this positive matrix by $s_\ell$. They are the $N$ positive roots, counted with multiplicity, of
\begin{equation}
    \det\!\left(
        \Omega_N^{-1/2}B_{\lambda,N}\Omega_N^{-1/2}-sI
    \right)
    =
    \frac{\det(B_{\lambda,N}-s\Omega_N)}{\det\Omega_N}
    =0\,.
\end{equation}
The $s_\ell$ are dimensionless. At $\lambda=0$, for example, all $s_\ell=1$, while the $\nu_m$ reduce to the original oscillator frequencies.

Let $R$ be the orthogonal matrix whose columns are the corresponding orthonormal eigenvectors, and define $\hat x^\prime=R^T\hat x$ and $\hat y^\prime=R^T\hat y$. Applying the same rotation to both quadratures preserves their canonical commutators, $[\hat x^\prime_\ell,\hat y^\prime_m]=i\delta_{\ell m}$. The original vacuum is invariant under this rotation because $x^Tx=x^{\prime T}x^\prime$. For the auxiliary ground state, the rotation diagonalises the exponent, so its wave-function factorises as
\begin{equation}
    \psi_{\lambda,N}(x^\prime)
    \propto
    \exp\!\left[-\frac12\sum_{\ell=1}^{N}s_\ell(x^\prime_\ell)^2\right] = \prod_{\ell=1}^{N} \exp\!\left[-\frac{s_\ell}{2}(x^\prime_\ell)^2\right]\,.
\end{equation}
For each mode, the probability density of $x^\prime_\ell$ is thus proportional to $\exp[-s_\ell(x^\prime_\ell)^2]$. Fourier transforming the corresponding wave-function gives a probability density proportional to $\exp[-(y^\prime_\ell)^2/s_\ell]$ for the conjugate quadrature $y^\prime_\ell$. Reading off the variances and defining the squeezing parameter by $r_\ell\coloneqq\tfrac12\log s_\ell$, we obtain
\begin{equation}
    \operatorname{Var}(\hat x^\prime_\ell) =\frac{1}{2s_\ell} =\frac12 e^{-2r_\ell}\,, \quad \operatorname{Var}(\hat y^\prime_\ell) =\frac{s_\ell}{2} =\frac12 e^{2r_\ell}\,.
\end{equation}
In this rotated basis, the auxiliary ground state is a product of single-mode squeezed vacua. In the original Fourier basis it is a multimode squeezed vacuum, generally with correlations between different modes. The squeezing parameters and the corresponding quadrature directions depend on the target and reference profiles through $d_N$.

\subsection{Relative variance and physical excitation energy}
Since $\hat\Phi_{f|A,w}=d_N^T\hat q$, its variance in the centred state follows by contracting the field covariance in \cref{eq:opt_covariances} with $d_N$. The auxiliary ground energy follows by summing the normal-mode zero-point energies and subtracting the original vacuum energy. This gives
\begin{equation}
    \begin{aligned}
        \sigma_N^2(\lambda)
        &= d_N^T\Gamma_{qq}^{(\lambda,N)}d_N
         = \frac{\hbar}{2}d_N^T B_{\lambda,N}^{-1}d_N\,,\\
        e_N(\lambda)
        &= \frac{\hbar}{2}\sum_{\ell=1}^{N}\nu_\ell
           - \frac{\hbar}{2}\sum_{j=1}^{N}\omega_j
         = \frac{\hbar}{2}\operatorname{Tr}
           \bigl(B_{\lambda,N}-\Omega_N\bigr)\,.
    \end{aligned}
    \label{eq:opt_finite_variance}
\end{equation}
The energy entering the preparation budget is the expectation of $\hat H_{\mathrm{exc}}$ in this same state. Substituting \cref{eq:opt_covariances} into its kinetic and potential terms gives
\begin{equation}
    \begin{aligned}
        E_N(\lambda)
        &= \frac12\operatorname{Tr}\!\left[
               \Gamma_{pp}^{(\lambda,N)}
               + \Omega_N^2\Gamma_{qq}^{(\lambda,N)}
           \right]
           - \frac{\hbar}{2}\operatorname{Tr}\Omega_N\\
        &= \frac{\hbar}{4}\operatorname{Tr}\!\left[
               B_{\lambda,N}
               + \Omega_N^2 B_{\lambda,N}^{-1}
               - 2\Omega_N
           \right]\,.
    \end{aligned}
    \label{eq:opt_finite_energy}
\end{equation}
Taking the expectation of \cref{eq:opt_auxiliary} relates the two energies,
\begin{equation}
    e_N(\lambda)
    = E_N(\lambda)+\frac{\lambda}{2}\sigma_N^2(\lambda)\,.
\end{equation}

\subsection{Resolvent representation and integral formulas}
\label{app:Resolvent_representation_and_integral_formulas}
At finite mode cutoff, spectral calculus and the scalar identity
\begin{equation}
    a^{-1/2} = \frac{2}{\pi}\int_0^\infty\frac{\dd\xi}{a+\xi^2}\,,
    \quad a>0\,,
\end{equation}
give
\begin{equation}
    K_{\lambda,N}^{-1/2}
    = \frac{2}{\pi}\int_0^\infty \dd\xi\,
      (K_{\lambda,N}+\xi^2 I)^{-1}\,.
    \label{eq:opt_inverse_square_root}
\end{equation}
Here $\xi$ is an auxiliary variable with dimensions of frequency. Define
\begin{equation}
    M_N(\xi)=\Omega_N^2+\xi^2 I\,,
    \quad
    G_N(\xi) = d_N^T M_N(\xi)^{-1}d_N = \sum_{j=1}^{N}\frac{d_j(w)^2}{\omega_j^2+\xi^2}\,.
    \label{eq:qrf-resolvent-mode-sum}
\end{equation}
Since $K_{\lambda,N}+\xi^2 I=M_N(\xi)+\lambda d_Nd_N^T$, the Sherman-Morrison formula~\cite{ShermanMorrison1950} yields
\begin{equation}
    (K_{\lambda,N}+\xi^2 I)^{-1} = M_N(\xi)^{-1} - \frac{\lambda\,M_N(\xi)^{-1}d_Nd_N^T M_N(\xi)^{-1}}{1+\lambda G_N(\xi)}\,.
    \label{eq:opt_rank_one_resolvent}
\end{equation}
The denominator is positive for $\lambda\geq0$. Contracting with $d_N$ gives
\begin{equation}
    \begin{aligned}
        d_N^T(K_{\lambda,N}+\xi^2 I)^{-1}d_N
        &= G_N(\xi)- \frac{\lambda G_N(\xi)^2}{1+\lambda G_N(\xi)}\\
        &= \frac{G_N(\xi)}{1+\lambda G_N(\xi)}\,.
    \end{aligned}
    \label{eq:qrf-scalar-perturbed-contraction}
\end{equation}
Substituting \cref{eq:opt_inverse_square_root} and \cref{eq:qrf-scalar-perturbed-contraction} into the variance in \cref{eq:opt_finite_variance}, we obtain
\begin{equation}
    \sigma_N^2(\lambda)
    = \frac{\hbar}{\pi}\int_0^\infty \dd\xi \, \frac{G_N(\xi)}{1+\lambda G_N(\xi)}\,.
    \label{eq:qrf-variance-integral}
\end{equation}
For fixed target and reference profiles, $G_N$ is independent of $\lambda$. The mode sum in \cref{eq:qrf-resolvent-mode-sum} can therefore be evaluated once and used to compute the variance for any choice of the multiplier. For the normalised auxiliary ground state, the Hellmann-Feynman identity gives
\begin{equation}
    e_N^\prime(\lambda) = \expval{\frac{\partial\hat H_{\lambda,N}}{\partial\lambda}}_{\Psi_{\lambda,N}} = \frac12 \expval{\hat\Phi_{f|A,w}^{\,2}}_{\Psi_{\lambda,N}} = \frac12 \sigma_N^2(\lambda)\,,
    \label{eq:opt_hellmann_feynman}
\end{equation}
where the last equality follows because the state is centred. Since $e_N(0)=0$, integration yields
\begin{equation}
    \begin{aligned}
        e_N(\lambda)
        &= \frac12\int_0^\lambda \dd s \,\sigma_N^2(s)\\
        &= \frac{\hbar}{2\pi}\int_0^\infty \dd\xi\, \int_0^\lambda \dd s\, \frac{G_N(\xi)}{1+sG_N(\xi)}\\
        &= \frac{\hbar}{2\pi}\int_0^\infty \dd\xi\, \log\!\bigl[1+\lambda G_N(\xi)\bigr]\,.
    \end{aligned}
    \label{eq:qrf-auxiliary-energy-integral}
\end{equation}
The integrand is non-negative, so the order of integration may be exchanged. Finally, using $E_N(\lambda)=e_N(\lambda)-\lambda \sigma_N^2(\lambda)/2$, we obtain the physical excitation energy
\begin{equation}
    E_N(\lambda)
    =
    \frac{\hbar}{2\pi}\int_0^\infty \dd\xi\,
    \left[
        \log\!\bigl(1+\lambda G_N(\xi)\bigr)
        - \frac{\lambda G_N(\xi)}{1+\lambda G_N(\xi)}
    \right]\,.
    \label{eq:qrf-excitation-energy-integral}
\end{equation}

Equations~\cref{eq:qrf-variance-integral} and \cref{eq:qrf-excitation-energy-integral} give the energy-variance relation in parametric form. For a prescribed excitation energy $E>0$ and a non-zero reading, solving $E_N(\lambda_{E,N})=E$ fixes the multiplier. The state \cref{eq:opt_wavefunction} at this multiplier attains the minimum relative variance $\sigma_N^2(\lambda_{E,N})$. To obtain the slope of the optimal energy-variance curve, we keep the target and reference profiles fixed and differentiate $E_N(\lambda)=e_N(\lambda)-\lambda \sigma_N^2(\lambda)/2$. Using the Hellmann-Feynman identity $e_N^\prime(\lambda)=\sigma_N^2(\lambda)/2$, we find
\begin{equation}
    E_N^\prime(\lambda) = e_N^\prime(\lambda)  -\frac12 \sigma_N^2(\lambda) -\frac{\lambda}{2}(\sigma_N^2)^\prime(\lambda)  = -\frac{\lambda}{2}(\sigma_N^2)^\prime(\lambda)\,.
\end{equation}
For a non-zero reading, differentiating the integral for the variance gives
\begin{equation}
    (\sigma_N^2)^\prime(\lambda) =-\frac{\hbar}{\pi}\int_0^\infty \dd\xi\, \frac{G_N(\xi)^2}{[1+\lambda G_N(\xi)]^2} <0\,.
\end{equation}
Hence $E_N^\prime(\lambda)>0$ for $\lambda>0$. Since the prescribed budget fixes $\lambda_{E,N}$ through $E_N(\lambda_{E,N})=E$, the chain rule gives
\begin{equation}
    \frac{\dd}{\dd E}\sigma_N^2(\lambda_{E,N}) =\frac{(\sigma_N^2)^\prime(\lambda_{E,N})}{E_N^\prime(\lambda_{E,N})} =-\frac{2}{\lambda_{E,N}}\,, \quad E>0\,.
\end{equation}
At the endpoint $E=0$, the multiplier is $\lambda=0$ and $B_{0,N}=\Omega_N$. All squeezing parameters therefore vanish, and the optimal state is the original field vacuum. 

We have displayed above, what is the optimal state and its relative variance when the field is restricted to $N$ real, non-zero Fourier modes. We now say a few words about what happens when the cutoff is removed. Starting from the fact that $h_w=f-r_{A,w}$ is smooth on the circle, it is square integrable. Hence, we can use Parseval's identity
\begin{equation}
    \sum_{j=1}^{\infty}d_j(w)^2=\|h_w\|_{L^2}^2\,, \quad \omega_j\geq\omega_*=\frac{2\pi v}{L}>0\,,
    \label{eq:cutoff_basic_bounds}
\end{equation}
we also have another bound which comes from the minimum reachable frequency for non-zero modes. These two properties play distinct roles. Square summability controls the total weight of the Fourier modes omitted at finite $N$. The lower bound on the frequencies controls the denominators that enter the covariances and the energy. 

By the bounds above, the sum $G_N(\xi)$ in \cref{eq:qrf-resolvent-mode-sum} converges pointwise to
\begin{equation}
G(\xi)=\sum_{j=1}^{\infty}\frac{d_j^2}{\omega_j^2+\xi^2},
    \qquad
    0\leq G_N(\xi)\leq G(\xi)
    \leq \frac{\|d\|^2}{\omega_*^2+\xi^2}.
\end{equation}
At fixed $\lambda$, this integrable bound allows us to replace $G_N$ by $G$ in \cref{eq:qrf-variance-integral,eq:qrf-auxiliary-energy-integral,eq:qrf-excitation-energy-integral} by dominated convergence. The full auxiliary Hamiltonian can be diagonalised directly in Fock space~\cite{MatsuzawaSasakiUsami2021}. Moreover, extending the finite-cutoff ground states by the vacuum of the omitted modes adds only a vanishing vacuum-tail contribution to their auxiliary energy, so their expectation of the full auxiliary Hamiltonian tends to its ground-state energy.  This then implies that these states converge in norm to it, up to a phase~\cite{MatsuzawaSasakiUsami2021}. 

Finally, We choose $\lambda_E$ by solving $E_w(\lambda_E)=E$, where $E_w(\lambda)=\lim_{N\to\infty}E_N(\lambda)$. The same argument presented in this appendix then shows that the ground state at $\lambda_E$ attains the minimum variance at budget $E$ in the full Fock space.

\end{document}